\documentclass[reprint, showpacs, showkeys, superscriptaddress, aps, pre, twocolumn, 10pt]{revtex4-1}

\usepackage[T1]{fontenc}
\usepackage[utf8]{inputenc}
\usepackage{amsmath}
\usepackage{amssymb}
\usepackage{enumitem}
\usepackage{graphicx}
\usepackage{refstyle}
\usepackage{multirow}
\usepackage[normalem]{ulem}
\newcommand\redout{\bgroup\markoverwith
{\textcolor{black}{\rule[.5ex]{5pt}{0.5pt}}}\ULon}
\usepackage[
  citecolor=blue,
  colorlinks,
  linkcolor=blue,
  urlcolor=blue,
]{hyperref}
\usepackage[dvipsnames]{xcolor}

\newcommand{\be}{\begin{equation}}
\newcommand{\ee}{\end{equation}} 
\newcommand{\bea}{\begin{eqnarray}}
\newcommand{\eea}{\end{eqnarray}}

\begin{document}
\title{Evolutionary Dynamics of Delayed Replicator-Mutator Equation: Limit Cycle and Cooperation}
\author{Sourabh Mittal}
\email{sourabhmittal19@gmail.com}
\affiliation{
 Department of Physics,
  Indian Institute of Technology Kanpur,
  Uttar Pradesh 208016, India
}
\author{Archan Mukhopadhyay}
\email{archan@iitk.ac.in}
\affiliation{
 Department of Physics,
  Indian Institute of Technology Kanpur,
  Uttar Pradesh 208016, India
}
\author{Sagar Chakraborty}
\email{sagarc@iitk.ac.in}
\affiliation{
  Department of Physics,
  Indian Institute of Technology Kanpur,
  Uttar Pradesh 208016, India
}
%
\begin{abstract}
 Game theory deals with strategic interactions among players and evolutionary game dynamics tracks the fate of the players' populations under selection. In this paper, we consider the replicator equation for two-player-two-strategy games involving cooperators and defectors. We modify the equation to include the effect of mutation and also delay that corresponds either to the delayed information about the population state or in realizing the effect of interaction among players. By focusing on the four exhaustive classes of symmetrical games---\textcolor{black}{the} Stag Hunt game, \textcolor{black}{the} Snowdrift game, \textcolor{black}{the} Prisoners' Dilemma game, and \textcolor{black}{the} Harmony game---we analytically and numerically analyze delayed replicator-mutator equation to find the explicit condition for the Hopf bifurcation bringing forth stable limit cycle. The existence of the asymptotically stable limit cycle imply the coexistence of the cooperators and the defectors; and in the games, where defection is a stable Nash strategy, a stable limit cycle does provide a mechanism for evolution of cooperation. We find that while mutation alone can never lead to oscillatory cooperation state in two-player-two-strategy games, the delay can change the scenario. On the other hand, there are situations when delay alone cannot lead to the Hopf bifurcation in the absence of mutation in the selection dynamics.
\end{abstract}

\keywords{Evolutionary Game Theory, Replicator Dynamics, Delay Differential Equation, Hopf Bifurcation}
\maketitle

\section{Introduction}
\label{sec:1}

\textcolor{black}{It is perplexing that cooperation~\citep{axelord1981science,axelrod1984book,bourke2011book} should be ubiquitously present in social, ecological, and biological systems in spite of the fact that selfish actions by an agent fetch it relatively more benefit. While a lot of progress has been reported in understanding this phenomenon, a unanimous consensus is still to be reached. One reason simply is that in the complex systems under consideration, comprehending every aspect of the phenomenon and trying to give one sweeping reason behind it is extremely challenging if not impossible. This is where analyzing simple stylized strategic games like the famous Prisoner's Dilemma game becomes useful. Such games allow for a drastically simplified version of the problem of the evolution of cooperation and present a transparent abstraction of the problem. Based on these games, the theory of evolutionary games~\citep{smith1982book,nowak2006book} has been providing insights into the dilemma of the evolution of cooperation.} 

\textcolor{black}{In its most simple form~\citep{osborne2009book}, the classical game theory assumes that two rational players---each equipped with two strategies (actions)---play against each other by using one strategy each simultaneously to get some consistently quantifiable payoff (profit or loss). The information about the profit or the loss for a player corresponding to every strategic interaction is written down as an element in a $2 \times 2$ matrix called the payoff matrix. Here, rationality means consistency in decision-making: each decision-making player chooses the best action in accordance with his/her set of preferences that is complete and transitive. One also assumes that the players have common knowledge of the rules of the game. The most celebrated prediction of the game theory is that in such one-shot games, theoretically, a potential outcome corresponds to the Nash equilibrium~\citep{nash1950pnas} that is a pair of strategies (one from each player's strategy set) such that no player can benefit by unilateral deviating from his/her equilibrium strategy.}

\textcolor{black}{The evolutionary game theory does away with the requirements of the rationality and the common knowledge since the biological entities it is concerned with need not be rational. The organisms become players and the strategies become synonymous with the phenotypes of the organisms. The elements of the payoff matrices denote---within the Darwinian paradigm---the fitnesses that are most conveniently interpreted as the numbers of offsprings. The fitness of a phenotype, thus, is defined as the average payoff that one individual of that phenotype gets in the population of organisms of different phenotypes. In the absence of rationality, while the concept of the Nash equilibrium becomes redundant in the evolutionary game theory, it turns out that the evolutionary stable strategy (ESS) must be a Nash equilibrium. ESS is a strategy that when adopted by the whole population, the host population becomes resilient against an infinitesimal fraction of mutants playing some alternative strategy. Since the evolution is essentially a dynamical process, many dynamical equations modelling the evolution are in vogue~\citep{hofbauer1998book}. One of the most investigated evolutionary dynamical equation is the replicator equation~\cite{taylor1978mb}. It is remarkable that its asymptotic dynamical outcomes are related to the underlying one-shot game's Nash equilibrium or ESS~\cite{cressman2014pnas} so that one may predict the dynamical outcome of the evolutionary dynamics by analyzing the game-theoretic equilibrium concepts of the corresponding game.}

Analysis of nonlinear dynamical equations modelling various aspects~\cite{melbinger2010prl, gomezgardees2012pre, juul2013pre,requejo2016pre, ermentrout2016pre, blokhuis2018prl, jiang2018pre, artiges2019pre, allahverdyan2019pre,barfuss2019pre,wang2019pre,yamamoto2019pre} of evolutionary games is an exciting modern interdisciplinary research area that encompasses problems from biophysics, mathematics, economics, and sociology. In its simplest form, {a} strategic interaction in a game is supposed to lead to realization of payoff/fitness instantaneously. While most well-investigated game dynamics assume such strategic interactions, there are some systems where the effect of {an} interaction may take some time to set in~\citep{stepan2009rsta,erneux2017chaos}. The presence of delay affecting the course of dynamics essentially means that the underlying \textcolor{black}{decision-making} process has a memory associated with it and is not Markovian in nature. It is well known that the effect of memory is capable of inducing cooperation in dynamical games~\citep{alonso-sanz2009ijbc, alonso-sanz2009chaos, wang2015biosystems}.

        There are quite a few studies on the effect of delay in selection dynamics in evolutionary game theory, specifically, replicator dynamics. Inclusion of delay due to information lag in \textcolor{black}{the} evaluation of fitness for simple two-player-two-strategy games in the continuous replicator dynamics~\citep{yi1997jtb} reveal that the conditions for ESS \textcolor{black}{are} independent of delay and stability of the interior fixed point corresponding to mixed ESS depends on delay. Actually, two different types of delay can be envisaged~\citep{alboszta2004jtb}: one corresponds to the information lag (delayed information about the population state) and the other to the delay in realizing the effect of interaction among players. The former one is called social delay and the latter one biological delay. It has been shown that the mixed ESS is asymptotically stable for \textcolor{black}{a} small social delay but loses stability (unlike the case of biological delay) when the delay increases beyond a threshold value. 

    It is, thus, not surprising that a particular model of delay (that includes the social delay as one specific case) has been shown to induce limit cycle behaviour in two-player-two-strategy replicator dynamics~\citep{wesson2016ijbc}. Also, the Hopf bifurcation  \textcolor{black}{leading to the} emergence of limit cycle has been observed in the replicator dynamics of $N$-person Stag Hunt game~\citep{moreira2012jtb}. It is further known~\citep{ijima2012jtb} that the condition, at which the stability of equilibrium points of replicator dynamics corresponding to two-player-two-strategy symmetric games changes, does not depend on the distribution of delay. The consequences of delay, in general, are quite nontrivial and complex. For discrete dynamics, the relationship between the information lag about the phenotypic distribution and stability of \textcolor{black}{the} interior fixed point is not monotonic~\citep{ijima2011mss}. The same study also shows that for smoothed best response dynamics in anti-coordination games, the interior fixed point is stable for low probability of delay and unstable for \textcolor{black}{the} large probability of delay.  In another work~\citep{burridge2017epjb} involving a population of finite agents with a specific memory length of past interactions and playing snowdrift game, one notes that the fixed points may become unstable and give way to limit cycles for large memory length. 

What is surprising is that almost all the investigations (including the aforementioned works) on the effect of delay on the selection dynamics \textcolor{black}{ignore} the complications due to \textcolor{black}{the} omnipresent phenomenon of mutation. While the conventional mutation can be of biological origin, any shift in the strategy of an agent---assumed to play only pure strategies---can be interpreted as mutation.
    We feel that it is of immediate pragmatic interest to study selection-mutation dynamics with delay. The mutations can be \textcolor{black}{categorized} into two types~\citep{toupo2015pre}: multiplicative mutation that stands for the error in replication mechanism during \textcolor{black}{the} birth of offspring and additive mutation that models mutation in adults. Replicator dynamics containing the multiplicative mutation has been studied in the context of the problem of grammar acquisition~\citep{komarova2001jtb} among other problems.  The effect of the additive mutation in the replicator dynamics for rock-paper-scissor game~\citep{mobilia2010jtb,toupo2015pre} and \textcolor{black}{the} repeated Prisoner's Dilemma game~\citep{toupo2014ijbc} has been studied to find the Hopf bifurcation and limit cycles therein. It should be realized that \textcolor{black}{the} existence of \textcolor{black}{a} limit cycle is equivalent to \textcolor{black}{the} presence of cooperation in the systems as is elaborated throughout this paper.

The combined effect of delay and mutation on the evolutionary dynamics being a relatively unaddressed problem, we address this issue using delayed replicator-mutator dynamics for \textcolor{black}{the} two-player-two-strategy game. We consider both the multiplicative and the additive mutations and so in the \textcolor{black}{two-dimensional} mutation parameter space, we find the region where \textcolor{black}{a} stable limit cycle emerges following the Hopf bifurcation. Our specific attention is on the four classes of games, \emph{viz.}, \textcolor{black}{the} Snowdrift (SD), \textcolor{black}{the} Stag Hunt (SH), \textcolor{black}{the} Prisoner's Dilemma (PD), and \textcolor{black}{the} Harmony (HG) which are traditionally studied to understand the evolution of cooperation. Specifically, the flow of the paper is as follows: First we discuss the games that model cooperation in Sec.~\ref{sec:2}. Then, we propose the models of delay in Sec.~\ref{sec:3}, followed by a discussion on the linear stability analysis and what happens to the evolution of cooperation in the presence of delay and mutation. We conclude our paper in Sec.~\ref{sec:6}.
\section{Replicator-Mutator Equation}
\label{sec:2}
\textcolor{black}{The} replicator equation is one of the most widely used models of selection dynamics in evolutionary games. For an unstructured infinite population consisting of $n$ (pheno-)types (pure strategies), we denote the frequency of $i$th type by $x_{i}$; of course, $\sum_{j=1}^{n}x_j=1$. Let the fitness or the expected payoff of $i$th type be $f_i$. The average fitness of the population thus is $\phi = \sum_{j=1}^{n}x_j f_j$. The probability of multiplicative mutation, i.e., the probability that some of the $j$th type offsprings are born from the $i$th type individual is given by $Q_{ij}$ ($\sum_{j=1}^{n}Q_{ij}=1$). Furthermore, we assume a constant rate $\mu$ of additive mutation, i.e., adults of certain type changing their strategy to another corresponding to some other type. The resulting replicator-mutator equation mathematically can be expressed as,
	\begin{eqnarray}
	\dot{x}_i = \sum\limits_{j=1}^{n} x_j f_j Q_{ji} - \phi x_i -\mu(nx_i-1).
	\label{eq:replicator-mutator}
	\end{eqnarray}
Here all the terms are evaluated at the same instant of time as there is no delay in the system and hence, we do not show time $t$ explicitly as the argument of the variables.

As mentioned in Sec.~\ref{sec:1}, we are interested in comprehending the effect of delay and mutation on the evolution of cooperation. The evolution of cooperation has long been intriguing researchers~\citep{axelord1981science,axelrod1984book,shalom1986jp,hilbe2018nature}. Interesting dilemmas result in simple  \textcolor{black}{one-shot} two-player-two-strategy games (like \textcolor{black}{the} PD) when individuals defect to play non-Pareto-optimal Nash equilibrium when mutual cooperation could have fetched them more payoff. If ``cooperate'' and ``defect'' are the only two strategies under consideration, one may divide all the games into four classes~\citep{nowak1992nature,fang2002gdn,doebeli2004science,kummerli2007prsb,Barker2017book,hilbe2018nature}, \emph{viz.}, \textcolor{black}{the} SH, \textcolor{black}{the} SD, \textcolor{black}{the} PD, and \textcolor{black}{the} HG based on the fact how the symmetric Nash equilibria are related to cooperate strategy. 
\begin{figure*}
\centering
\includegraphics[scale=0.41]{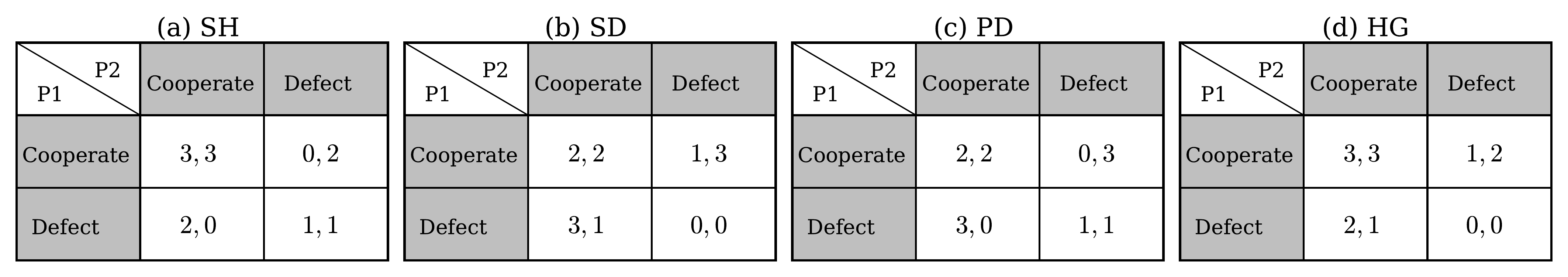} 
\caption{Presented are the typical payoff matrices of (a) \textcolor{black}{the} Stag Hunt game, (b) \textcolor{black}{the} Snowdrift game, (c) \textcolor{black}{the} Prisoners' dilemma game, and (d) \textcolor{black}{the} Harmony  game, that have been explicitly used for the calculations in this paper. P1 and P2 refer to the two players playing the games.}  
\label{fig:1}
\end{figure*}

In other words, consider that the normal bimatrix form of \textcolor{black}{one-shot}  symmetric two-player-two-strategy game is represented as follows:
 \begin{eqnarray*}  
\centering
\begin{tabular}{cc|c|c|}
		& \multicolumn{1}{c}{} & \multicolumn{2}{c}{\textbf{Player $2$}}\\
		& \multicolumn{1}{c}{} & \multicolumn{1}{c}{Cooperate} & \multicolumn{1}{c}{\,\,\,\,Defect\,\,\,\,}\\\cline{3-4} 
		\multirow{2}*{\textbf{Player $1$}} & Cooperate & $a,a$ & $b,c$ \\\cline{3-4}
		& Defect & $c,b$ & $d,d$ \\\cline{3-4} 
\end{tabular}
\end{eqnarray*}
where \textcolor{black}{the first element in each cell is  the payoff of player $1$ and the second element is that of player $2$.} \textcolor{black}{Payoff elements} $a$, $b$, $c$, and $d$ are real numbers. The ordinal relationship between the payoff elements define the aforementioned four classes of games:
\begin{enumerate}
\item[SH:] $c<a$ and $a>d>b$. This \textcolor{black}{one-shot} game corresponds to two symmetric Nash equilibria---cooperate and defect, and one mixed symmetric Nash equilibrium. The essence of \textcolor{black}{the} SH (coordination) game is conveniently exemplified~\cite{skyrms2001paapa} as follows: given that hunting down a stag (largest payoff) requires cooperation between two players, a non-cooperating player can only catch a hare (smaller payoff) while the other player, being alone in the chase of the stag, returns \textcolor{black}{empty-handed} (least payoff).
\item[SD:] $c>a$ and $b>d$. This \textcolor{black}{one-shot} game corresponds to one symmetric Nash equilibrium in which the players randomize their strategies. The anti-coordination SD game appears in the scenario where two individuals are trapped in a big snowdrift that blocks a road. Either individual has the strategy to either cooperate in clearing the blockage or to wait for the other to clear it. Of course, a \textcolor{black}{free-rider}  (defector) has the most advantage but there is the risk that if both keep waiting for the other to clear the blockage, then they both incur \textcolor{black}{a} maximum loss by being stuck forever.
\item[PD:]  $c>a>d>b$. Mutual defection is the unique Nash equilibrium in the corresponding \textcolor{black}{one-shot} game which probably is the most famous one in the popular literature. The dilemma showcased in the game is that although (cooperate, cooperate) strategy profile is Pareto-optimal, the (symmetric) Nash equilibrium corresponds to mutual defection.
 \item[HG:] $c<a$ and $b>d$. Mutual cooperation is the unique Nash equilibrium of this game.
\end{enumerate}

In this context, since the folk theorems~\cite{cressman2014pnas} connect the point attractors of replicator equation to the corresponding Nash equilibrium, an understanding of the selection dynamics is paramount. Being concerned with two-player-two-strategy games only, $n=2$. Let the fraction of cooperators be $x$, i.e., $x_1=x$. This implies that the fraction of defectors is $1-x$, i.e., $x_2=1-x$. Also, the fitness of $i$th type is \textcolor{black}{$f_i =\sum_{j=1}^{n}{\Pi}_{ij}x_j$}, where
 \begin{eqnarray}
\begin{tabular}{c}
 ${\sf \Pi}$ =
 $\begin{bmatrix}  
a  & b   \\  c & d \\ 
\end{bmatrix}$.
\end{tabular}\vspace{2mm}
\label{eq:PayOff_A}
\end{eqnarray}
Assuming that the fraction of accurate replication for both the types are same, i.e., $Q_{11}=Q_{22}=q\le1$, Eq.~(\ref{eq:replicator-mutator}) can be written as,
\begin{eqnarray}
\dot{x} &=& -x^3 [a-b-c+d] \nonumber \quad\\
&& + x^2[q(a-b+c-d)-2c+3d-b]\nonumber\quad\\
&&+ x[q(b-c+2d)+c-3d-2\mu] +d(1-q) +\mu.\quad
\label{eq:nodelay_explicit_form}
\end{eqnarray}
\textcolor{black}{The mutation matrix $\sf Q$, being symmetric and row stochastic, is completely specified by the single parameter, $q$}. On putting $q=1$ and $\mu=0$, we reach the case of no mutation, i.e., replicator dynamics. The replicator dynamics can have only one interior fixed point ($x^{*}=x_m$) along with two boundary fixed points ($x^{*}=0$ and $x^{*}=1$). Presence of mutation (either $q \in [0,1)$ or $\mu \in (0,1]$ or both) shifts the fixed points to, say, $X_{m}$, $X_{-}$, and $X_+$ respectively. To have a better understanding of the effect of the mutation parameters on the nature of the fixed points for the mentioned four classes of game, it is helpful to fix some appropriate numerical values for the payoff matrix elements as they facilitate analytically tractable calculations. The chosen payoff matrices are shown in Fig.~\ref{fig:1}. Unless otherwise specified, we henceforth exclusively work with these payoff matrices.
\begin{figure*}
	\centering
	\includegraphics[scale=0.42]{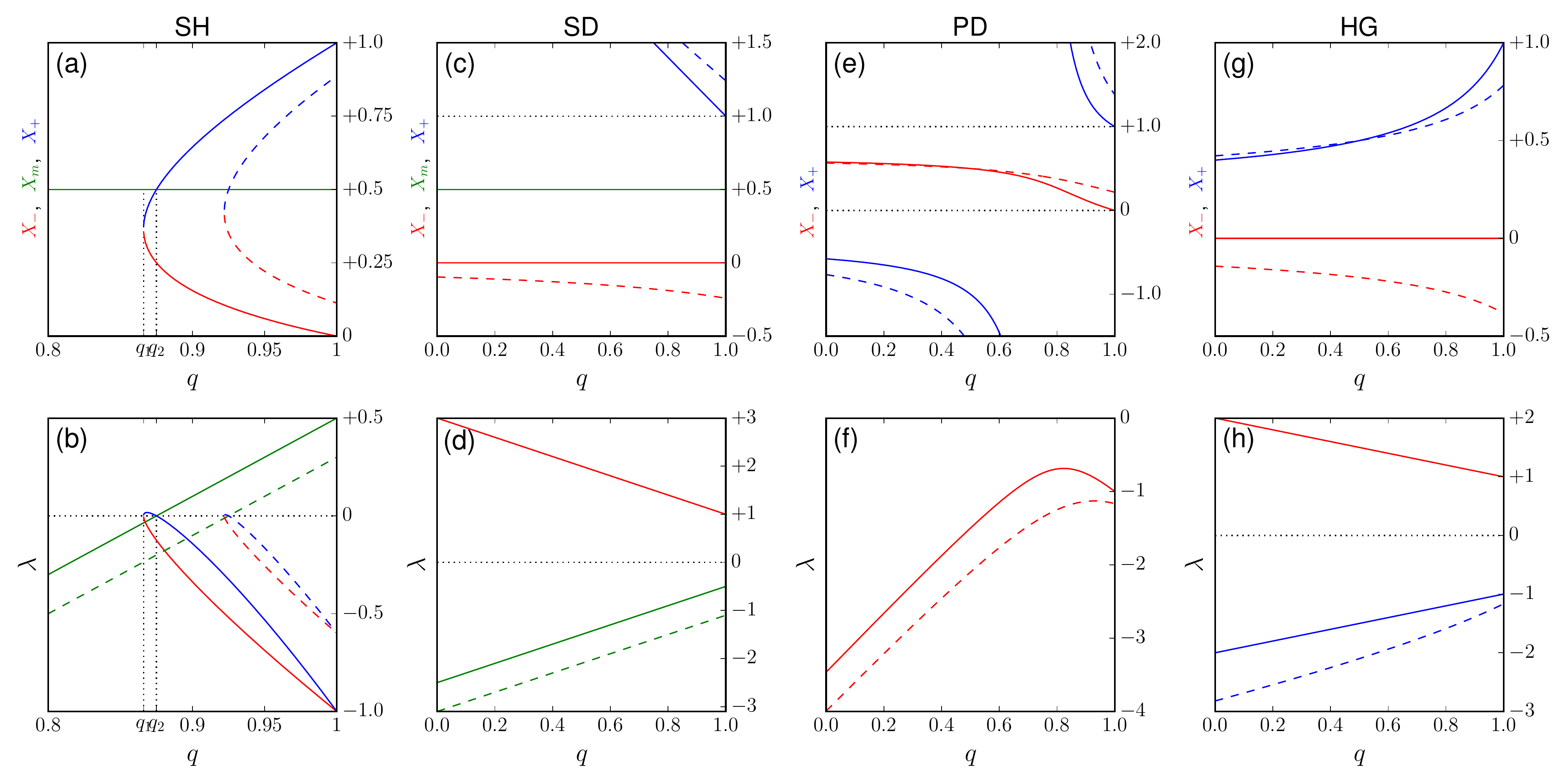}      		  
	\caption{\emph{(Colour online)} Linear stability of the fixed points of the replicator-mutator equation. In \textcolor{black}{the} top row depicts the variation of fixed points $X_m$ (green), $X_-$ (red) and $X_+$ (blue) with the change in mutation parameter $q$ (for two values of parameter $\mu$) when the payoff matrix corresponds to (a) \textcolor{black}{the} SH game, (c) \textcolor{black}{the} SD game, (e) \textcolor{black}{the} PD game, and (g) \textcolor{black}{the} HG. The corresponding eigenvalue, $\lambda$, of the Jacobian found in the course of linear stability analysis is plotted in the bottom row (following the colour conventions used for the top row) for (b) \textcolor{black}{the} SH game, (d) \textcolor{black}{the} SD game, (f) \textcolor{black}{the} PD game, and (h) \textcolor{black}{the} HG. \textcolor{black}{The} solid line stands for $\mu=0$ while the dashed line is for a non-zero $\mu$, specifically, $0.1$ for the SH game and $0.3$ for the other three games.}\label{fig:2}
\end{figure*}

In the case of \textcolor{black}{the} SH game, the fixed points of the dynamics are
\begin{eqnarray*}
&&X_m = \frac{1}{2},\\
&&X_- = \frac{1}{2}\left(-\sqrt{4q^2-4\mu-3}+2q-1\right),\\
&&X_+ = \frac{1}{2} \left(\sqrt{4q^2-4\mu-3}+2q-1\right),
 \label{eq:FP_Stag-hunt}
\end{eqnarray*}
as shown in Fig.~\ref{fig:2}(a).
Fixed points $X_-$ and $X_+$ are real if $q\ge q_1={(\sqrt{3+4\mu})}/{2}$. This means that $X_{+}$ and $X_{-}$ can exist only when $0 \leq \mu \leq {1}/{4}$. It can also be noted that transcritical bifurcation occurs (Fig.~\ref{fig:2}(b)) at $q =q_2= ({7+4\mu})/{8}$ at which $X_+$ collides with $X_m$ and their natures of stability get interchanged.
 
The fixed points of the replicator-mutator dynamics for \textcolor{black}{the} SD game are
\begin{eqnarray*}
&&X_m = \frac{1}{2},\\
&&X_- = \frac{1}{2} \left(-\sqrt{4q^2-12q+4\mu+9}-2q+3\right),\\
&&X_+ = \frac{1}{2} \left(\sqrt{4q^2-12q+4\mu+9}-2q+3\right).
 \label{eq:FP_Snow-Drift}
\end{eqnarray*}
In presence of mutation, the fixed point $X_+$ is unphysical (i.e., $X_+\notin [0,1]$). The other fixed point $X_-$ is unphysical whenever $\mu \ne 0$ (irrespective of the value of $q$). When $\mu=0$, $X_-=0$ for all possible values of $q$. The fixed points and their stabilities are depicted in Fig.~\ref{fig:2}(c)-(d).

On solving Eq.~(\ref{eq:nodelay_explicit_form}) for \textcolor{black}{the} PD game, following fixed points---as illustrated in Fig.~\ref{fig:2}(e)-(f)---are obtained:
\begin{eqnarray*}
&&X_- = \frac{-\sqrt{17q^2 -12q\mu -28q+4\mu^2 +12\mu +12}+q+2\mu}{2(4q-3)},\\
&&X_+ = \frac{\sqrt{17q^2 -12q\mu -28q+4\mu^2 +12\mu +12}+q+2\mu}{2(4q-3)}.
 \label{eq:FP_PD}
\end{eqnarray*}
The fixed point $X_+$ is unphysical when mutation of either kind is present. In the absence of mutation $X_+$ is physical and attains the value $X_+=1$. The fixed point $X_{-}$ always exists as an interior fixed point in \textcolor{black}{the} presence of mutation. In \textcolor{black}{the} absence of any mutation, $X_-=0$. One may note that mutation makes coexistence of cooperators and defectors possible in \textcolor{black}{the} PD game dynamics.

Out of the four classes of games that we are considering, \textcolor{black}{the} HG is the only one where the cooperate-strategy turns out to be a dominant strategy and hence pure a Nash equilibrium. The fixed points of its replicator-mutator dynamics are given as:
\begin{eqnarray*}
&&X_- = \frac{-\sqrt{ (2-q-2\mu)^2 -4\mu(4q-5)}+q+2\mu -2}{2(4q-5)},\\
&&X_+ = \frac{ \sqrt{ (2-q-2\mu)^2 -4\mu(4q-5)}+q+2\mu -2}{2(4q-5)}.
 \label{eq:FP_Harmony}
\end{eqnarray*}
 The fixed point $X_{-}$ is unphysical in \textcolor{black}{the} presence of $\mu$-mutation. When $\mu=0$, for any possible $q$, the fixed point $X_-=0$. So, only $X_+$ is the fixed point of practical importance. We exhibit all the possible fixed points and their linear stabilities in Fig.~\ref{fig:2}(g)-(h).

In all the four \textcolor{black}{classes} of games, as expected, there is no limit cycle behaviour as the phase space is one-dimensional for the corresponding autonomous replicator-mutator equation. While an interior stable fixed point (like in \textcolor{black}{the} PD game) does imply \textcolor{black}{the} emergence of cooperation, \textcolor{black}{the} existence of \textcolor{black}{a} stable limit cycle provides another mechanism for the establishment of cooperation. In general, \textcolor{black}{the} inclusion of delay in the replicator-mutator equation makes the phase space effectively \textcolor{black}{infinite-dimensional} and hence there is \textcolor{black}{a} possibility of limit cycles. So, \emph{does delay induce cooperation in the games in the presence of mutation? How does the interplay between mutation and delay affect the dynamics?} These are the main questions that we now seek to address in the rest of this paper.

\section{Delayed Replicator-Mutator Equation}
\label{sec:3}
 A glance at the replicator-mutator equation for two-player-two-strategy games involving cooperators and defectors suggests that---whether delay corresponds to the delayed information about the population state or in realizing the effect of interaction among players---mathematically, delay has to appear in either the state, ${\bf x}$, or the expected fitnesses, ${f_i}\, (i\in\{1,2\})$. Thus, to be very general, we speak of a doublet $(\tau_1,\tau_2)$, where $\tau_1$ and $\tau_2$ are respectively the characteristics delays corresponding to the cooperators and the defectors. How the delays are incorporated in the dynamics is a different issue that we describe in what immediately follows.
\subsection{Two Types of Delay: Social and Biological}	
\label{sec:3.1}
Consider the rather general case of an infinite unstructured population consisting of $n$ types of individuals. If the information regarding the fitnesses in the population is delayed by $\tau_i$ (social delay) for each type, or in other words, if  individuals use past information about the population to evaluate their fitnesses, then the replicator-mutator dynamics given by  Eq.~(\ref{eq:replicator-mutator}) can be written as,
	\begin{eqnarray}
	\dot{x}_i(t) = \sum\limits_{j=1}^{n} x_j(t) f_j(t-\tau_j) Q_{ji} - \phi x_i(t) -\mu\big[nx_i(t)-1\big],\qquad
	\label{eq:RM-soc}
	\end{eqnarray}
where $\phi = \sum_{j=1}^{n} x_j(t) f_j(t-\tau_j)$. If we use the payoff matrix form given by Eq.~(\ref{eq:PayOff_A}), the replicator-mutator dynamics with social delay takes the explicit form given below:
\begin{eqnarray}
\dot{x} &=& -x^2 {x_{\tau_{1}}} [a-b] -x^2 {x_{\tau_{2}}} [d-c]+ x{x_{\tau_{1}}}[q(a-b)]\nonumber\\
&& + x{x_{\tau_{2}}}[q(c-d)-2c+2d]+x^2[d-b]\nonumber\qquad\\
&&+ {x_{\tau_{2}}}[c-d+q(d-c)] + x[q(b+d)-2d-2 \mu] \nonumber\qquad\\
&& +d(1-q) +\mu~,
\label{eq:social_explicit_form}
\end{eqnarray}
where the subscripts denote the respective delay in the corresponding arguments, \emph{e.g.}, $x_{\tau_1}$ means $x(t-\tau_1)$. For the case of no mutation and $\tau_1=\tau_2$ (symmetric delay), this model has been introduced ~\cite{alboszta2004jtb}. The case of asymmetric delay ($\tau_1\ne\tau_2$) case has also been studied~\cite{tembine2007article} in the absence of mutation.

The second type of delay---biological delay---that interests us comes into action in the systems where the effect of an interaction is not instantaneous and consequently there is a delay in realizing the payoff of an interaction. Thus, both the fitness of an individual and the state of the population used in replicator-mutator dynamics should be calculated at the past instant when the interaction happened. Mathematically, the delayed replicator-mutator dynamics should be cast as,
\begin{eqnarray}
\dot{x}_i(t) = \sum_{j=1}^{n} x_j(t-\tau_j)f_j(t-\tau_j) Q_{ji} -\phi x_i-\mu\big[nx_i(t)-1\big] ,\quad
\label{eq:RM-bio}
\end{eqnarray}
where $\phi = \sum_{j=1}^{n} f_j(t-\tau_j) x_j(t-\tau_j)$. Again, for payoff matrix given by Eq.~(\ref{eq:PayOff_A}), the replicator-mutator dynamics with biological delay has the following form:
 \begin{eqnarray}
  \dot{x}& =& -x x_{\tau_{1}}^2 [a-b] -x x_{\tau_{2}}^2 [d-c] \nonumber\\
 &&+ x{x_{\tau_{1}}}[-b]+ x{x_{\tau_{2}}}[-c+2d] +x_{\tau_{1}}^2 [q(a-b)] \nonumber\\
&& +x_{\tau_{2}}^2 [d-c-q(d-c)]+ {x_{\tau_{1}}}[bq]  \nonumber\\
&&+{x_{\tau_{2}}}[(c-2d)(1-q)]+ x[-d-2 \mu] +d(1-q) +\mu .\quad\,\,
\label{eq:bio_explicit_form}
 \end{eqnarray}
This has also been studied~\cite{alboszta2004jtb} in the case of symmetric delay and no mutation.
 
Equipped with the aforementioned governing equations, we want to attempt \textcolor{black}{to understand} the combined effect of mutation and delay on the evolution of cooperation. Specifically, in what follows, we work with symmetric delay $(\tau_1=\tau_2=\tau)$ and two types of asymmetric delay--- $(0,\tau)$ and $(\tau,0)$. This choice helps us to reduce the number of delay parameters to work with only one, i.e., $\tau$.
 
\subsection{Linear Stability Analysis}
\label{sec:4}
The next logical step in the search of \textcolor{black}stable limit cycle in Eq.~(\ref{eq:social_explicit_form}) and Eq.~(\ref{eq:bio_explicit_form}) is to perform linear stability analyses on the equations and look out for the Hopf bifurcation. The eigenvalues ($\lambda$) dictating the stability of the corresponding fixed point can be obtained from the characteristic equation~\citep{roose2007book}:
\begin{equation}
\lambda + \alpha e^{-\lambda \tau} = \beta,
\label{eq:characteristic_eq_3}
\end{equation}
where $\alpha$ and $\beta$ are real functions of the system parameters, {\emph{viz.}}, payoff matrix elements and mutation parameters. \textcolor{black}{The parameters $\alpha$ and $\beta$ can be expressed in terms of the Jacobians, $J_0 = (d\dot{x}/dx)_{x=x^*}$ and $J_{\tau} = (d\dot{x}/dx_{\tau})_{x=x^*}$. Explicitly, $\alpha$ and $\beta$ are $-J_{\tau}$ and $J_0$ respectively.} 

The infinite possible solutions ~\citep{asl2003jdsmc, shinozaki2006automatica} to Eq.~(\ref{eq:characteristic_eq_3}) can be written as 
\begin{equation}
\lambda_k = \beta + \frac{W_k(-\alpha\tau e^{-\beta\tau})}{\tau},\, \forall k \in \mathbb Z;
\label{eq:solution_lambda}
\end{equation}
where $W_k$ is the $k$th branch of the Lambert $W$ function. \textcolor{black}{The principal branch of the Lambert $W$ function corresponds to $k=0$}.

 The use of the Lambert $W$ function in the analysis of stability of fixed points in delay differential equations is widely discussed in standard literature~\citep{asl2003jdsmc,yu2017jmci, shinozaki2006automatica, yi2007mbe,lehtonen2016mee,bortz2015ifac}. $\operatorname{Re}(\lambda_k)$---real part of $\lambda_k$---is maximum for $k=0$ as $W_0$ has maximum real part among all $W_k$~\citep{yi2007mbe,shinozaki2006automatica}. Thus, one can say that the stability of a fixed point in the presence of delay is solely determined by the eigenvalue, $\lambda_0$, corresponding to the principal branch of the Lambert $W$ function. We expect emergence of stable limit cycle as a consequence of Hopf bifurcation about a fixed point when it has purely imaginary $\lambda_0$ and all other eigenvalues are such that $\operatorname{Re}(\lambda_k)<0$~\citep{kuang1993book,gopalsamy1992book, arino2006book,yu2016namc,li2008csf}. 
 
 Consequently, we focus on the solution for $\lambda_0$. We first note that $\lambda_0$ is real if $W_0$ is real. This implies $-\alpha\tau e^{-\beta\tau} \geq -1/e$ or $\alpha \leq{e^{\beta \tau}}/{e\tau}$. Therefore, if $\alpha \leq [e^{\beta \tau}/e\tau]\, \forall \tau > 0$---or in other words, $\alpha \leq \min_\tau[e^{\beta \tau}/e\tau]$---then $\lambda_0$ is always real $\forall\tau >0$. If, however, $\alpha > \min_\tau[e^{\beta \tau}/e\tau]$, then $\lambda_0$ is complex for some $\tau >0$. It is easy to note that condition $\alpha > \min_\tau[e^{\beta \tau}/e\tau]$ is equivalent to condition $\alpha>\max\{0,\beta\}$.

Now, consider that $\alpha>\max\{0,\beta\}$ and assume that $\exists\,\tau = \tau_H>0$ at which $\lambda _0= i \omega$, a purely imaginary number. By putting this in Eq.~(\ref{eq:characteristic_eq_3}), and separating the real and the imaginary parts, we get,
\begin{eqnarray} 
&&\omega^2 = \alpha^2 - \beta^2, \\
&&\tau_H = \frac{1}{\sqrt{\alpha^2 - \beta^2}}\cos^{-1}\left(\frac{\beta}{\alpha}\right).
    \label{eq:cond_imag}
\end{eqnarray}
For $\tau_H$ to be real, $\omega$ must be real; in this case it means that in addition to $\alpha>\beta$ (trivially satisfied because of the condition: $\alpha>\max\{0,\beta\}$), $\alpha+\beta>0$ must also hold. These considerations give us the recipe to find the stable limit cycle solutions in the two-player-two-strategy replicator-mutator equation with social and biological delays, either symmetric or asymmetric: All one has to do is to find the system parameters such that the corresponding $\alpha$ and $\beta$ obey both the inequalities---$\alpha>\max\{0,\beta\}$ and $\alpha+\beta>0$.

\subsection{Limit Cycle and Cooperation}
\label{sec:5}

 \begin{figure*}
 \centering
 \includegraphics[scale=0.42]{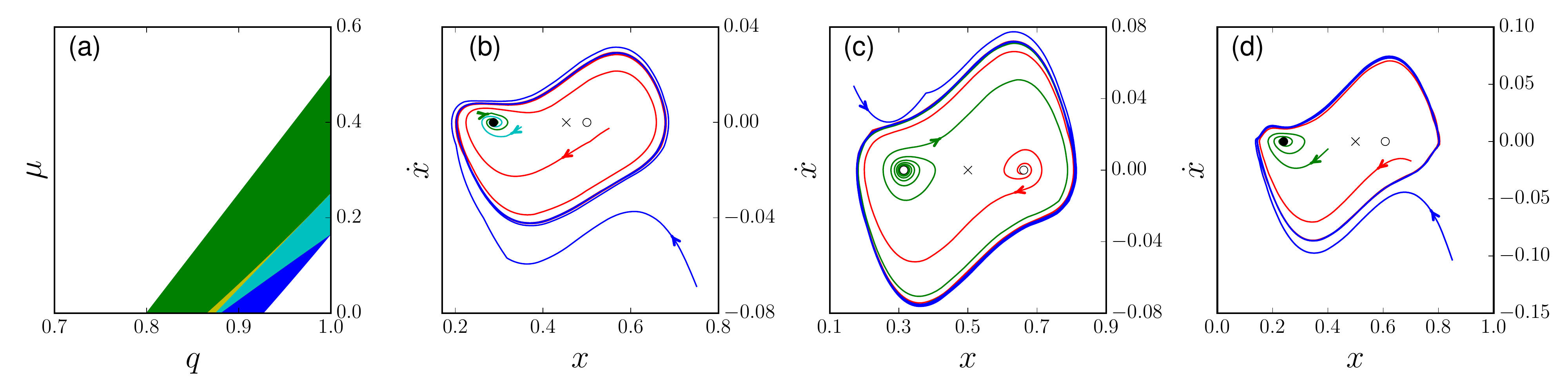}      		  
 \caption{\emph{(Colour online)} Stable limit cycle emerges following the Hopf bifurcation in the SH game for social asymmetric delay $(0,\tau)$: Colour-filled areas in subplot (a) marks the total region in the mutation parameter space where stable limit cycle emerges at some threshold value of delay. The fixed points that undergo the Hopf bifurcation in the green, the yellow, the cyan, and the blue colours are $X_m$, $X_m$ and $X_-$, $X_-$ and $X_+$, and $X_+$ respectively. Subplots (b), (c), and (d) showcase the phase diagram corresponding to some point belonging to the green, the cyan, and the blue region respectively in $q$-$\mu$ space. In particular, we fix $q$, $\mu$, and $\tau$ as $0.87,\,0$, and $9.1$; $0.988,\,0.196$, and $6.5$; and $0.924,\,0.07$, and $12$ respectively in subplots (b), (c), and (d). The filled circle, unfilled circle, and cross in the phase plots respectively represent stable focus, unstable focus, and saddle. The red, the blue, and the green curves are representative phase trajectories approaching attractors.}
 \label{fig:4}
 \end{figure*} 
\begin{figure}
\centering
\includegraphics[scale=0.42]{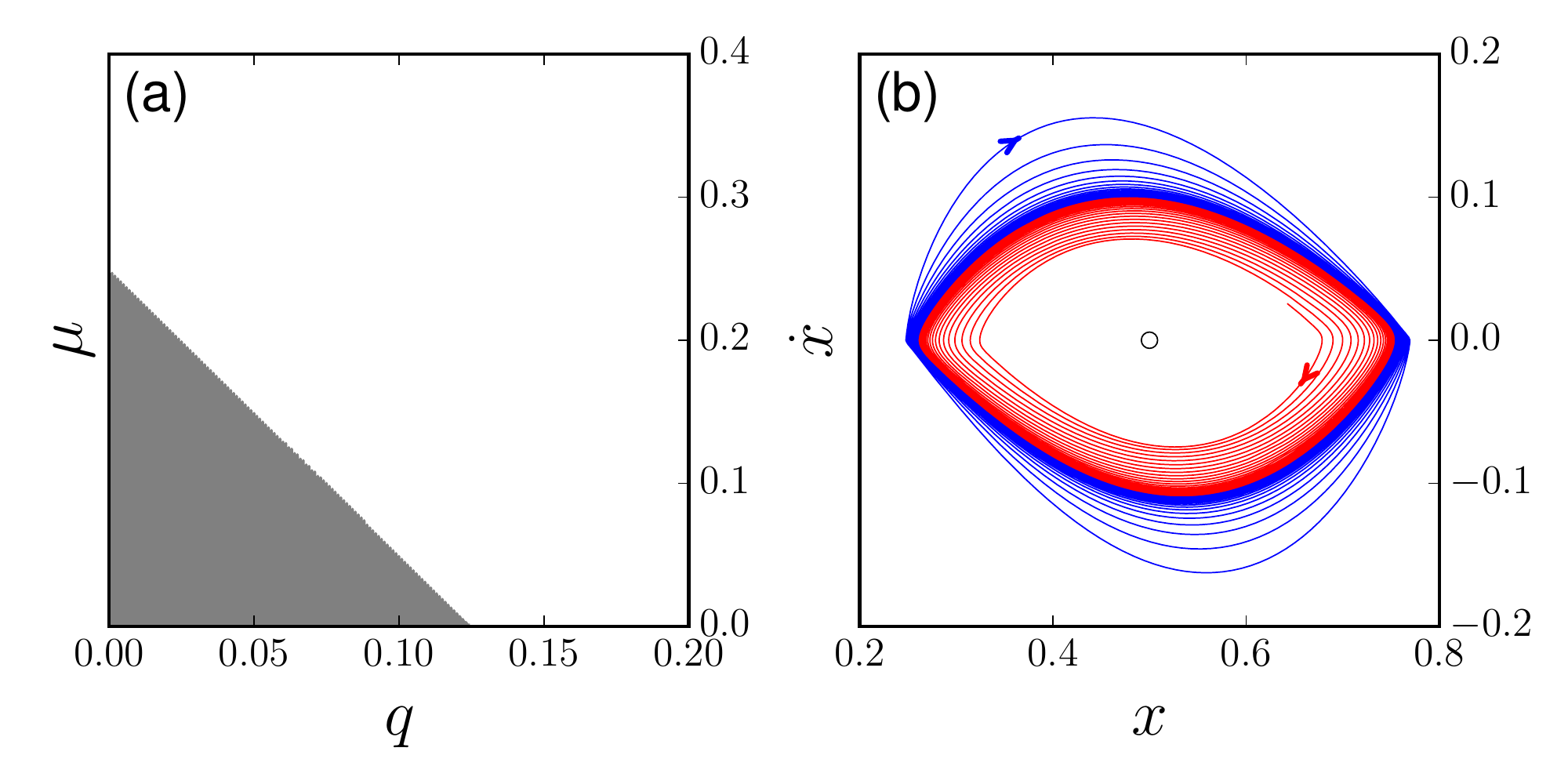}  
\caption{\emph{(Colour online)} Stable limit cycle emerges following the Hopf bifurcation in the SH game for biological symmetric delay $(\tau,\tau)$: Grey area in subplot (a) marks the region in the mutation parameter space where stable limit cycle emerges at some threshold value of delay. The fixed point that undergoes the Hopf bifurcation is $X_m$. Subplot (b) showcases an illustrative phase diagram corresponding to $q=0.097$, $\mu=0.001$, and $\tau=10$ picked from the grey region. The unfilled circle represents unstable focus, $X_m$; and the red and the blue curves are representative phase trajectories approaching the limit cycle from inside and outside respectively.}
\label{fig:5}
\end{figure}
\begin{figure}
\centering
\includegraphics[scale=0.37]{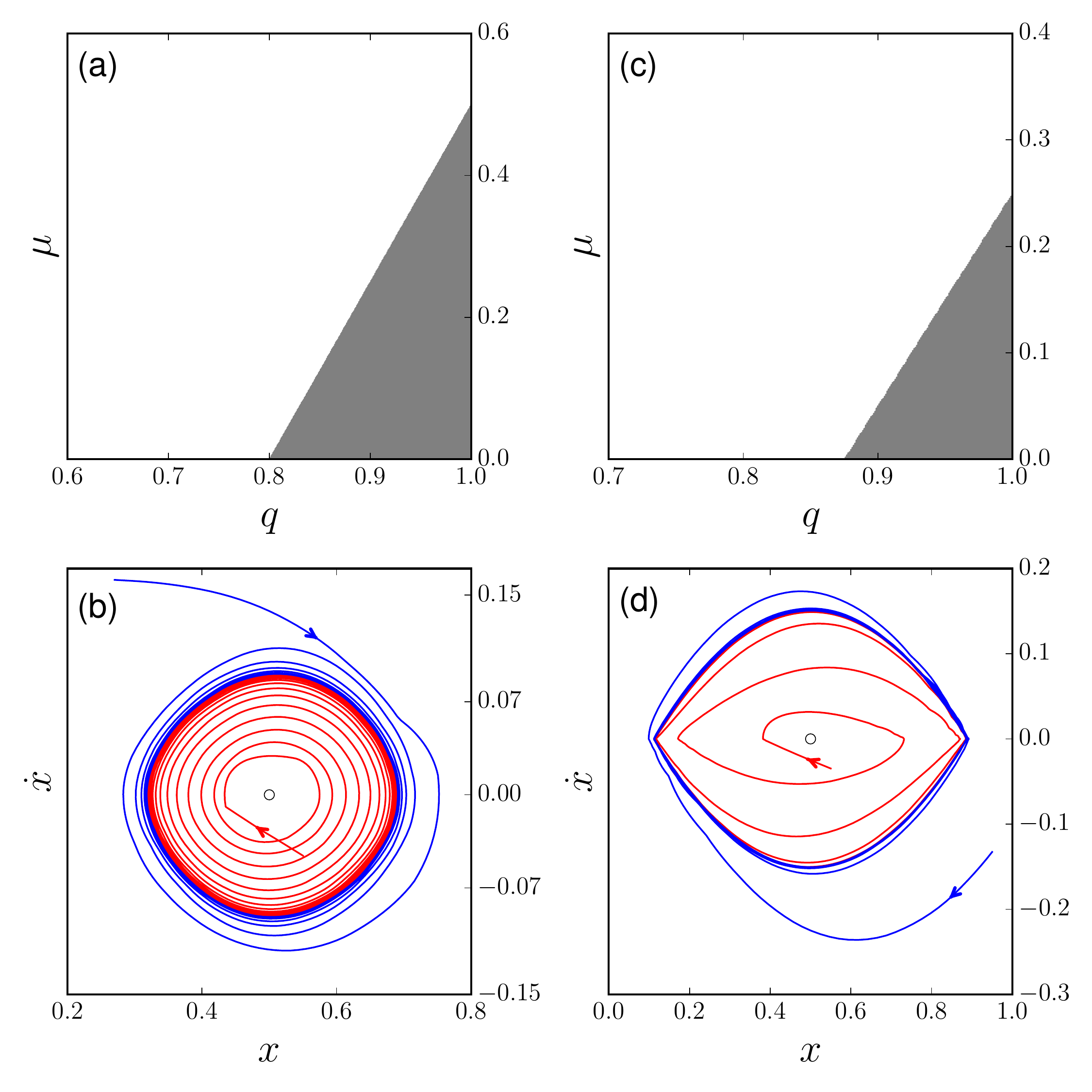}  
\caption{\emph{(Colour online)} Stable limit cycle emerges following the Hopf bifurcation in the SD game for social delay: Grey areas in subplot (a) and (c) mark the regions in the mutation parameter space where stable limit cycle emerges at some threshold value of asymmetric delay $(0,\tau)$ and symmetric delay $(\tau,\tau)$ respectively. The fixed point that undergoes the Hopf bifurcation is $X_m$. Subplot (b) and (d) respectively showcases illustrative phase diagrams corresponding to $q=0.9$, $\mu=0.1$, and $(\tau_1,\tau_2)=(0,\tau)=(0,5)$; and  $q=1$, $\mu=0.1$ and $(\tau_1,\tau_2)=(\tau,\tau)=(30,30)$ picked from the grey regions. The unfilled circle represent unstable focus, $X_m$; and the red and the blue curves are representative phase trajectories approaching the limit cycles from inside and outside respectively.}
\label{fig:6}
\end{figure}
\begin{figure}
\centering
\includegraphics[scale=0.42]{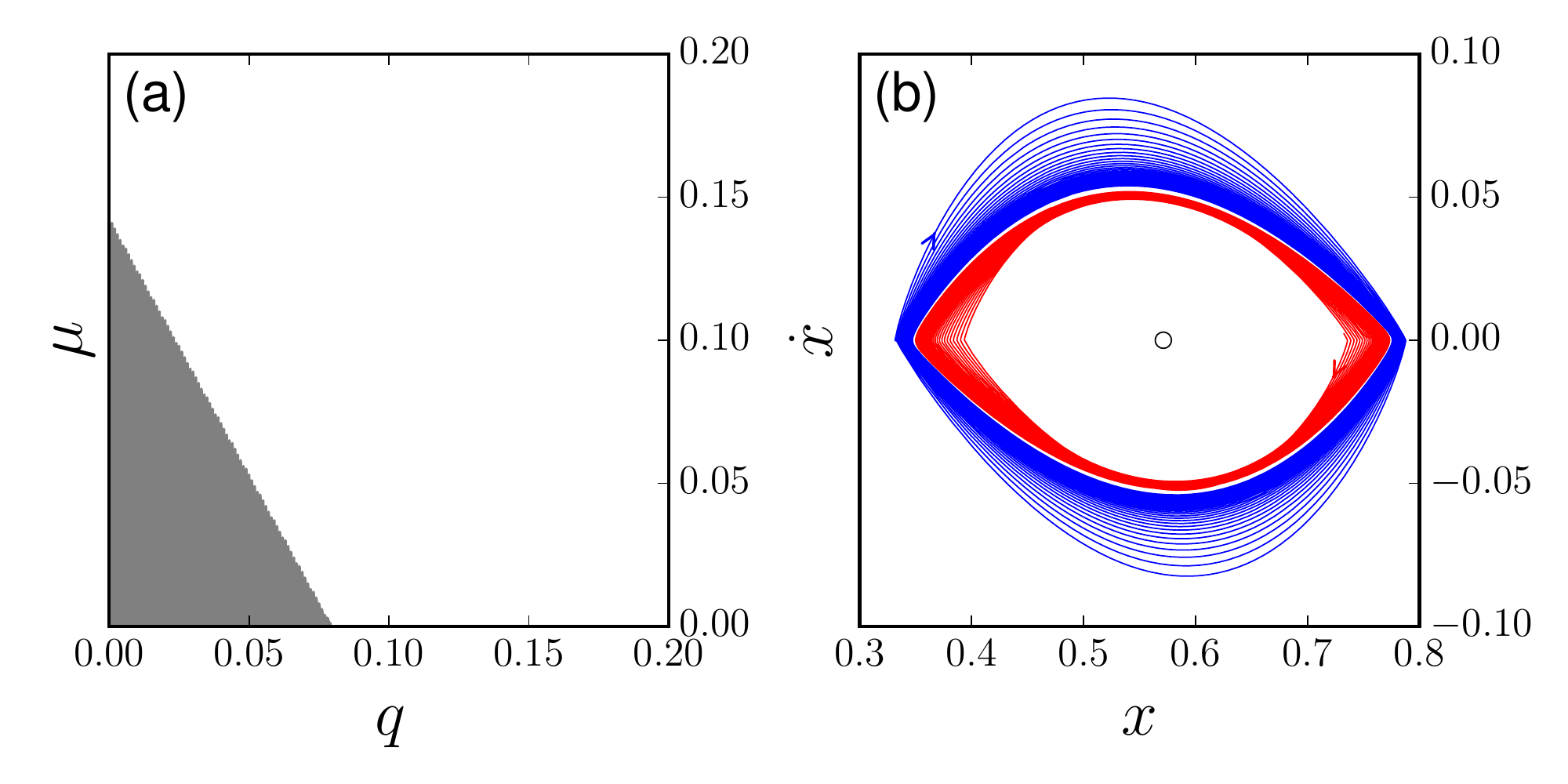}      		  
\caption{\emph{(Colour online)} \textcolor{black}{A} stable limit cycle emerges following the Hopf bifurcation in the PD game for biological symmetric delay $(\tau,\tau)$: Grey area in subplot (a) marks the region in the mutation parameter space where stable limit cycle emerges at some threshold value of delay. The fixed point that undergo the Hopf bifurcation is $X_-$. Subplot (b) showcases an illustrative phase diagram corresponding to $q=0.07$, $\mu=0$, and $\tau=20$ picked from the grey region. The unfilled circle represent unstable focus, $X_-$; and the red and the blue curves are representative phase trajectories approaching the limit cycle from inside and outside respectively.}
\label{fig:7}
\end{figure}
 \begin{figure}
\centering
\includegraphics[scale=0.42]{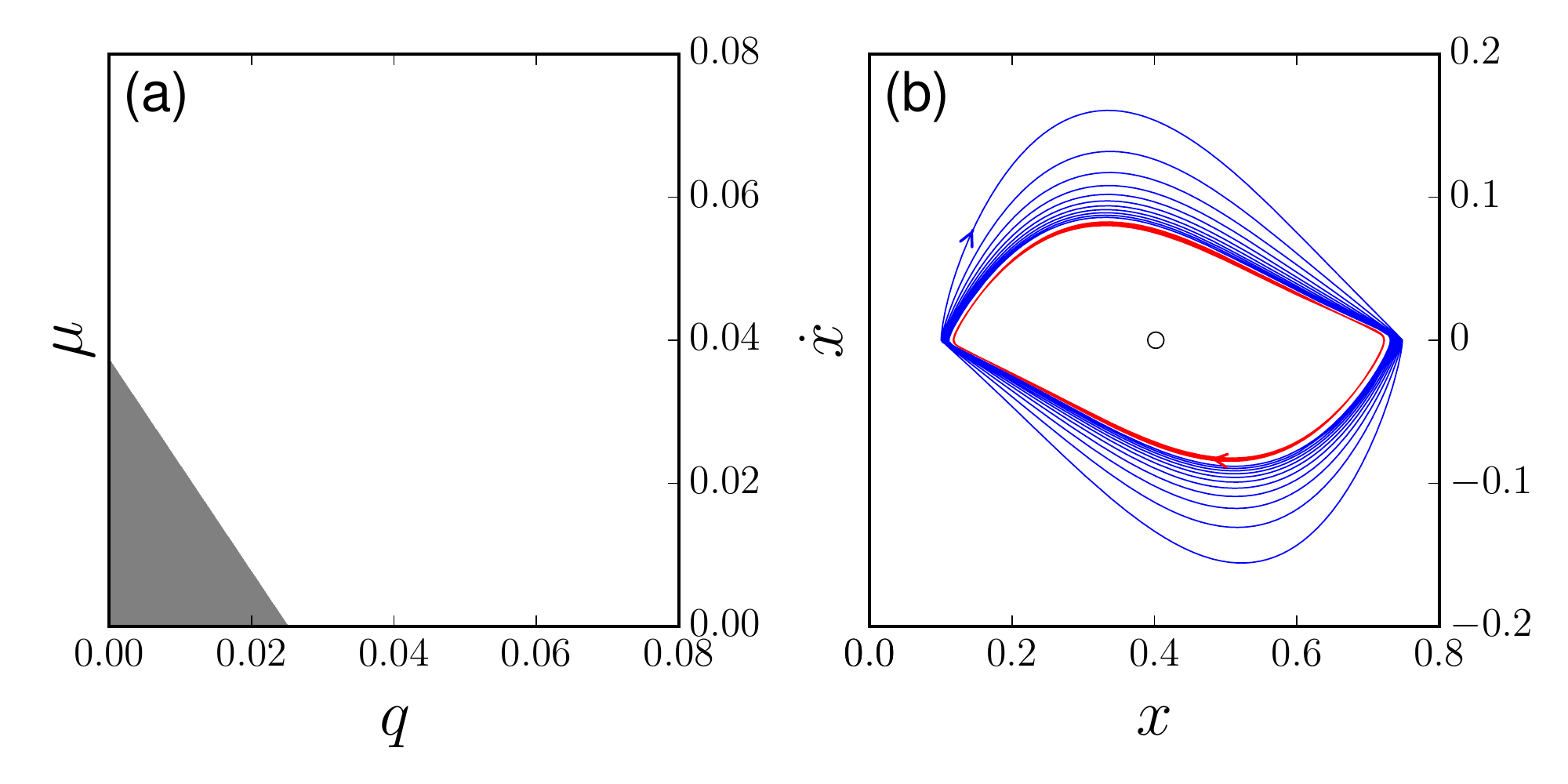}      		  
\caption{\emph{(Colour online)} Stable limit cycle emerges following the Hopf bifurcation in the HG for biological asymmetric delay $(\tau,0)$: Grey area in subplot (a) marks the region in the mutation parameter space where \textcolor{black}{a} stable limit cycle emerges at some threshold value of delay. The fixed point that undergoes the Hopf bifurcation is $X_+$. Subplot (b) showcases an illustrative phase diagram corresponding to $q=0.014$, $\mu=0.002$, and $\tau=20$ picked from the grey region. The filled circle represents unstable focus, $X_+$; and the red and the blue curves are representative phase trajectories approaching the limit cycle from inside and outside respectively.}
\label{fig:8}
 \end{figure}
 Since \textcolor{black}{the} existence of a stable limit cycle in the dynamics means coexistence of cooperators and defectors, the level of cooperation in the game is closely associated with limit cycles. This is especially important in \textcolor{black}{those} cases where defect strategy corresponds to an attracting fixed point of the dynamics in the absence of mutation and delay that when present induce cooperation in the games through limit cycles. Having already fixed the payoff matrix elements (see Fig.~\ref{fig:1}), it is the set of mutation parameters, $q$ and $\mu$, which decide when stable limit cycles are possible in accordance with the recipe outlined in the immediately preceding subsection. After investigating all the four classes of games for both the symmetric and the asymmetric delays of both the social and the biological types, below we report only the cases when \textcolor{black}{a} stable limit cycle could be firmly established analytically and numerically. Interested readers may refer Appendix~\ref{app:A} to find the summary of what happens when there is no mutation but delay is in play.

To begin with, we consider the SH game. We find that all the three fixed points ($X_m$, $X_-$, and $X_+$) undergo the Hopf bifurcation leading to the emergence of \textcolor{black}{a} stable limit cycle for social asymmetric delay where delay is only in the defector's fitness (refer Fig.~\ref{fig:4}). It is clear from the figure that delay alone, in the absence of any mutation, cannot lead to any limit cycle. A stable limit cycle is also possible in the case of symmetric biological delay when the fixed point $X_m$ undergoes Hopf bifurcation as illustrated in Fig.~\ref{fig:5}). 

Recall that when mutation is present, $X_m$ is the only physical fixed point of the replicator-mutator dynamics for the SD game. This fixed point changes stability and gives way to the Hopf bifurcation when social delay---whether symmetric or asymmetric (delay only in defector's fitness)---is in play. The illustrative limit cycles and the corresponding region of mutation parameter space are presented in Fig.~\ref{fig:6}. As an aside, we point out that unlike the SH game, in the absence of any mutation, i.e.,  $q=1$ and $\mu=0$, we can find stable limit cycle~\cite{wesson2016ijbc} in the SD game.

The game of \textcolor{black}{the} PD is different from both the SH and the SD games in the sense that it allows \textcolor{black}{the emergence of a} stable limit cycle when symmetric biological delay is incorporated in the replicator-mutator equation (see Fig.~\ref{fig:7}). The Harmony game is similar to the PD game in the sense that it also requires biological delay for exhibiting limit cycle behaviour, however, unlike the PD game, the biological delay has to be asymmetric such that the delay is only in the cooperator's fitness (see Fig.~\ref{fig:8}). $X_-$ and  $X_+$ undergo the Hopf bifurcation in the PD game and the HG respectively.
\section{Discussion and Conclusion}
\label{sec:6}
Before we conclude, let us first succinctly point out the salient features of aforementioned scenarios of the Hopf bifurcation: Firstly, it is interesting to note that while the stable limit cycles appear at relatively low values of additive mutation, it is not always so with the multiplicative mutation; for a limit cycle to appear in the case of social delay, the multiplicative mutation has to be relatively weak (high $q$-values). Whenever a limit cycle is born in the presence of biological delay, the multiplicative mutation is needed to be quite high (low $q$-values). Secondly, out of all the cases studied, the HG is the only case where the delay has to be solely in the fitness of the cooperators for limit cycles to exist and it should be recalled that in the HG, cooperate-strategy is the sole dominant Nash strategy (hence, ESS) unlike the other three classes of games. Thirdly, in the presence of  optimal delay, the SH game, the PD game, or the HG can possess stable limit cycles only if there is \textcolor{black}{a} finite non-zero mutation in the system. Fourthly, it is worth noting that the SH is the only one game where \textcolor{black}{a} stable limit cycle emerges for both social delay and biological delay, and the SD is the only game where biological delay doesn't lead to a stable limit cycle.

We remind the readers that to arrive at the results listed above, we have modified the replicator-mutator dynamics corresponding to two-player-two-strategy games---in particular, \textcolor{black}{the} SH, \textcolor{black}{the} SD, \textcolor{black}{the} PD, and  \textcolor{black}{the} HG---to include the social and the biological delays. To the best of our knowledge, the insightful interplay between the mutation (both additive and multiplicative) and the delays has never been investigated either analytically or numerically as undertaken in this paper. It is obvious that the search for \textcolor{black}{a} limit cycle in such systems is linked with the bigger question of \textcolor{black}{the} evolution of cooperation through simple but instructive games. Also, seen from another perspective, the delay and the mutation leads to coexistence of (pheno-)types in a population.

Nevertheless, an important question may and should be raised: Two-player-two-strategy symmetric games can be classified into twelve ordinal classes~\cite{pandit2018chaos} that can be collected into four game-types---\textcolor{black}{the} SH, \textcolor{black}{the} SD, \textcolor{black}{the} PD, and \textcolor{black}{the} HG---based on how cooperate strategy fares or, equivalently, what the Nash equilibria are. While the PD game has only one type of ordinal game, the SH, the SD, and the HG has respectively three, three, and five ordinally inequivalent payoff matrices and corresponding games. Moreover within each ordinally distinct game, there can be games that are connected by some cardinal transformation. In simpler terms, while this paper deals with two free parameters, $q$ and $\mu$, actually even the payoff matrix elements could be treated as parameters. So, how would all the results qualitatively and quantitatively change if even the payoff matrix elements were varied? One immediately notes that a study spanning the full \textcolor{black}{six-dimensional} parameter space would be daunting. Nevertheless, as an indication of what to expect, we present a detailed analysis of the three ordinally inequivalent SD games and \textcolor{black}{the} cardinally related SD games in Appendix~\ref{app:B} and Appendix~\ref{app:C} respectively.

It is worth pointing out that there are many possible future research directions that can be pursued following this paper: We have limited our analysis only to two-player-two-strategy games because our intention has been to bring the intricacy of the delayed dynamics to the fore and for this purpose, \textcolor{black}{these} games are the simplest yet conveniently non-trivial. Extension of our work to the games with more strategies---\emph{e.g.}, the rock-paper-scissors game~\cite{weissing1991gem} and \textcolor{black}{the} problem of grammar acquisition~\cite{komarova2001jtb}---would definitely be exciting. It is known that the replicator equation with additive mutation \textcolor{black}{leads} to coexistence in the game~\cite{toupo2015pre} but how the game behaves under multiplicative mutations and delay remains unexplored. Following the route of investigation delineated herein one should be able to attack similar problems in the discrete replicator equations~\cite{pandit2018chaos}, in the repeated games, and also in other types of selection-dynamics~\cite{hofbauer1998book}. Furthermore, one could start with a model of selection-replication-mutation dynamics in a finite population and study the effect of the delay therein with a view to contrasting the resulting stochastic dynamics with the corresponding deterministic dynamics obtained in the limit of infinite populations. Last but not the least, the simultaneous effects of the delay and the mutations in \textcolor{black}{a} structured population may spring a few surprises.
 \bibliography{Mittal_etal_manuscript.bib}

\begin{thebibliography}{63}%
\makeatletter
\providecommand \@ifxundefined [1]{%
 \@ifx{#1\undefined}
}%
\providecommand \@ifnum [1]{%
 \ifnum #1\expandafter \@firstoftwo
 \else \expandafter \@secondoftwo
 \fi
}%
\providecommand \@ifx [1]{%
 \ifx #1\expandafter \@firstoftwo
 \else \expandafter \@secondoftwo
 \fi
}%
\providecommand \natexlab [1]{#1}%
\providecommand \enquote  [1]{``#1''}%
\providecommand \bibnamefont  [1]{#1}%
\providecommand \bibfnamefont [1]{#1}%
\providecommand \citenamefont [1]{#1}%
\providecommand \href@noop [0]{\@secondoftwo}%
\providecommand \href [0]{\begingroup \@sanitize@url \@href}%
\providecommand \@href[1]{\@@startlink{#1}\@@href}%
\providecommand \@@href[1]{\endgroup#1\@@endlink}%
\providecommand \@sanitize@url [0]{\catcode `\\12\catcode `\$12\catcode
  `\&12\catcode `\#12\catcode `\^12\catcode `\_12\catcode `\%12\relax}%
\providecommand \@@startlink[1]{}%
\providecommand \@@endlink[0]{}%
\providecommand \url  [0]{\begingroup\@sanitize@url \@url }%
\providecommand \@url [1]{\endgroup\@href {#1}{\urlprefix }}%
\providecommand \urlprefix  [0]{URL }%
\providecommand \Eprint [0]{\href }%
\providecommand \doibase [0]{http://dx.doi.org/}%
\providecommand \selectlanguage [0]{\@gobble}%
\providecommand \bibinfo  [0]{\@secondoftwo}%
\providecommand \bibfield  [0]{\@secondoftwo}%
\providecommand \translation [1]{[#1]}%
\providecommand \BibitemOpen [0]{}%
\providecommand \bibitemStop [0]{}%
\providecommand \bibitemNoStop [0]{.\EOS\space}%
\providecommand \EOS [0]{\spacefactor3000\relax}%
\providecommand \BibitemShut  [1]{\csname bibitem#1\endcsname}%
\let\auto@bib@innerbib\@empty
\bibitem [{\citenamefont {Axelrod}\ and\ \citenamefont
  {Hamilton}(1981)}]{axelord1981science}%
  \BibitemOpen
  \bibfield  {author} {\bibinfo {author} {\bibfnamefont {R.}~\bibnamefont
  {Axelrod}}\ and\ \bibinfo {author} {\bibfnamefont {W.~D.}\ \bibnamefont
  {Hamilton}},\ }\href {http://www.jstor.org/stable/1685895} {\bibfield
  {journal} {\bibinfo  {journal} {Science}\ }\textbf {\bibinfo {volume}
  {211}},\ \bibinfo {pages} {1390} (\bibinfo {year} {1981})}\BibitemShut
  {NoStop}%
\bibitem [{\citenamefont {Axelrod}(1984)}]{axelrod1984book}%
  \BibitemOpen
  \bibfield  {author} {\bibinfo {author} {\bibfnamefont {R.}~\bibnamefont
  {Axelrod}},\ }\href {https://books.google.co.in/books?id=NJZBCGbNs98C} {\emph
  {\bibinfo {title} {The Evolution of Cooperation}}},\ Basic books\ (\bibinfo
  {publisher} {Basic Books},\ \bibinfo {year} {1984})\BibitemShut {NoStop}%
\bibitem [{\citenamefont {Bourke}(2011)}]{bourke2011book}%
  \BibitemOpen
  \bibfield  {author} {\bibinfo {author} {\bibfnamefont {A.}~\bibnamefont
  {Bourke}},\ }\href {https://books.google.co.in/books?id=\_pWCN6iDydgC} {\emph
  {\bibinfo {title} {Principles of Social Evolution}}},\ Oxford Series in
  Ecology and Evolution\ (\bibinfo  {publisher} {Oxford University Press},\
  \bibinfo {year} {2011})\BibitemShut {NoStop}%
\bibitem [{\citenamefont {Smith}(1982)}]{smith1982book}%
  \BibitemOpen
  \bibfield  {author} {\bibinfo {author} {\bibfnamefont {J.~M.}\ \bibnamefont
  {Smith}},\ }\href@noop {} {\emph {\bibinfo {title} {Evolution and the Theory
  of Games}}}\ (\bibinfo  {publisher} {Cambridge University Press},\ \bibinfo
  {address} {Cambridge},\ \bibinfo {year} {1982})\BibitemShut {NoStop}%
\bibitem [{\citenamefont {Nowak}(2006)}]{nowak2006book}%
  \BibitemOpen
  \bibfield  {author} {\bibinfo {author} {\bibfnamefont {M.~A.}\ \bibnamefont
  {Nowak}},\ }\href@noop {} {\emph {\bibinfo {title} {Evolutionary Dynamics:
  Exploring the Equations of Life}}}\ (\bibinfo  {publisher} {Harvard
  University Press},\ \bibinfo {year} {2006})\BibitemShut {NoStop}%
\bibitem [{\citenamefont {Osborne}(2009)}]{osborne2009book}%
  \BibitemOpen
  \bibfield  {author} {\bibinfo {author} {\bibfnamefont {M.~J.}\ \bibnamefont
  {Osborne}},\ }\href {https://ideas.repec.org/b/oxp/obooks/9780195322484.html}
  {\emph {\bibinfo {title} {{Introduction to Game Theory: International
  Edition}}}}\ (\bibinfo  {publisher} {Oxford University Press},\ \bibinfo
  {year} {2009})\BibitemShut {NoStop}%
\bibitem [{\citenamefont {Nash}(1950)}]{nash1950pnas}%
  \BibitemOpen
  \bibfield  {author} {\bibinfo {author} {\bibfnamefont {J.~F.}\ \bibnamefont
  {Nash}},\ }\href {\doibase 10.1073/pnas.36.1.48} {\bibfield  {journal}
  {\bibinfo  {journal} {Proc. Natl. Acad. Sci.}\ }\textbf {\bibinfo {volume}
  {36}},\ \bibinfo {pages} {48} (\bibinfo {year} {1950})}\BibitemShut {NoStop}%
\bibitem [{\citenamefont {Hofbauer}\ and\ \citenamefont
  {Sigmund}(1998)}]{hofbauer1998book}%
  \BibitemOpen
  \bibfield  {author} {\bibinfo {author} {\bibfnamefont {J.}~\bibnamefont
  {Hofbauer}}\ and\ \bibinfo {author} {\bibfnamefont {K.}~\bibnamefont
  {Sigmund}},\ }\href
  {https://www.amazon.com/Evolutionary-Games-Population-Dynamics-Hofbauer-ebook/dp/B01LZ8VDVA%3FSubscriptionId%3D0JYN1NVW651KCA56C102%26tag%3Dtechkie-20%26linkCode%3Dxm2%26camp%3D2025%26creative%3D165953%26creativeASIN%3DB01LZ8VDVA}
  {\emph {\bibinfo {title} {Evolutionary Games and Population Dynamics}}}\
  (\bibinfo  {publisher} {Cambridge University Press},\ \bibinfo {address}
  {Cambridge},\ \bibinfo {year} {1998})\BibitemShut {NoStop}%
\bibitem [{\citenamefont {Taylor}\ and\ \citenamefont
  {Jonker}(1978)}]{taylor1978mb}%
  \BibitemOpen
  \bibfield  {author} {\bibinfo {author} {\bibfnamefont {P.~D.}\ \bibnamefont
  {Taylor}}\ and\ \bibinfo {author} {\bibfnamefont {L.~B.}\ \bibnamefont
  {Jonker}},\ }\href {\doibase 10.1016/0025-5564(78)90077-9} {\bibfield
  {journal} {\bibinfo  {journal} {Math. Biosci.}\ }\textbf {\bibinfo {volume}
  {40}},\ \bibinfo {pages} {145} (\bibinfo {year} {1978})}\BibitemShut
  {NoStop}%
\bibitem [{\citenamefont {Cressman}\ and\ \citenamefont
  {Tao}(2014)}]{cressman2014pnas}%
  \BibitemOpen
  \bibfield  {author} {\bibinfo {author} {\bibfnamefont {R.}~\bibnamefont
  {Cressman}}\ and\ \bibinfo {author} {\bibfnamefont {Y.}~\bibnamefont {Tao}},\
  }\href {\doibase 10.1073/pnas.1400823111} {\bibfield  {journal} {\bibinfo
  {journal} {Proc. Natl. Acad. Sci.}\ }\textbf {\bibinfo {volume} {111}},\
  \bibinfo {pages} {10810} (\bibinfo {year} {2014})}\BibitemShut {NoStop}%
\bibitem [{\citenamefont {Melbinger}\ \emph {et~al.}(2010)\citenamefont
  {Melbinger}, \citenamefont {Cremer},\ and\ \citenamefont
  {Frey}}]{melbinger2010prl}%
  \BibitemOpen
  \bibfield  {author} {\bibinfo {author} {\bibfnamefont {A.}~\bibnamefont
  {Melbinger}}, \bibinfo {author} {\bibfnamefont {J.}~\bibnamefont {Cremer}}, \
  and\ \bibinfo {author} {\bibfnamefont {E.}~\bibnamefont {Frey}},\ }\href
  {\doibase 10.1103/physrevlett.105.178101} {\bibfield  {journal} {\bibinfo
  {journal} {Phys. Rev. Lett.}\ }\textbf {\bibinfo {volume} {105}} (\bibinfo
  {year} {2010}),\ 10.1103/physrevlett.105.178101}\BibitemShut {NoStop}%
\bibitem [{\citenamefont {G{\'{o}}mez-Garde{\~{n}}es}\ \emph
  {et~al.}(2012)\citenamefont {G{\'{o}}mez-Garde{\~{n}}es}, \citenamefont
  {Gracia-L{\'{a}}zaro}, \citenamefont {Flor{\'{\i}}a},\ and\ \citenamefont
  {Moreno}}]{gomezgardees2012pre}%
  \BibitemOpen
  \bibfield  {author} {\bibinfo {author} {\bibfnamefont {J.}~\bibnamefont
  {G{\'{o}}mez-Garde{\~{n}}es}}, \bibinfo {author} {\bibfnamefont
  {C.}~\bibnamefont {Gracia-L{\'{a}}zaro}}, \bibinfo {author} {\bibfnamefont
  {L.~M.}\ \bibnamefont {Flor{\'{\i}}a}}, \ and\ \bibinfo {author}
  {\bibfnamefont {Y.}~\bibnamefont {Moreno}},\ }\href {\doibase
  10.1103/physreve.86.056113} {\bibfield  {journal} {\bibinfo  {journal} {Phys.
  Rev. E}\ }\textbf {\bibinfo {volume} {86}} (\bibinfo {year} {2012}),\
  10.1103/physreve.86.056113}\BibitemShut {NoStop}%
\bibitem [{\citenamefont {Juul}\ \emph {et~al.}(2013)\citenamefont {Juul},
  \citenamefont {Kianercy}, \citenamefont {Bernhardsson},\ and\ \citenamefont
  {Pigolotti}}]{juul2013pre}%
  \BibitemOpen
  \bibfield  {author} {\bibinfo {author} {\bibfnamefont {J.}~\bibnamefont
  {Juul}}, \bibinfo {author} {\bibfnamefont {A.}~\bibnamefont {Kianercy}},
  \bibinfo {author} {\bibfnamefont {S.}~\bibnamefont {Bernhardsson}}, \ and\
  \bibinfo {author} {\bibfnamefont {S.}~\bibnamefont {Pigolotti}},\ }\href
  {\doibase 10.1103/physreve.88.022806} {\bibfield  {journal} {\bibinfo
  {journal} {Phys. Rev. E}\ }\textbf {\bibinfo {volume} {88}} (\bibinfo {year}
  {2013}),\ 10.1103/physreve.88.022806}\BibitemShut {NoStop}%
\bibitem [{\citenamefont {Requejo}\ and\ \citenamefont
  {D{\'{\i}}az-Guilera}(2016)}]{requejo2016pre}%
  \BibitemOpen
  \bibfield  {author} {\bibinfo {author} {\bibfnamefont {R.~J.}\ \bibnamefont
  {Requejo}}\ and\ \bibinfo {author} {\bibfnamefont {A.}~\bibnamefont
  {D{\'{\i}}az-Guilera}},\ }\href {\doibase 10.1103/physreve.94.022301}
  {\bibfield  {journal} {\bibinfo  {journal} {Phys. Rev. E}\ }\textbf {\bibinfo
  {volume} {94}} (\bibinfo {year} {2016}),\
  10.1103/physreve.94.022301}\BibitemShut {NoStop}%
\bibitem [{\citenamefont {Ermentrout}\ \emph {et~al.}(2016)\citenamefont
  {Ermentrout}, \citenamefont {Griffin},\ and\ \citenamefont
  {Belmonte}}]{ermentrout2016pre}%
  \BibitemOpen
  \bibfield  {author} {\bibinfo {author} {\bibfnamefont {G.~B.}\ \bibnamefont
  {Ermentrout}}, \bibinfo {author} {\bibfnamefont {C.}~\bibnamefont {Griffin}},
  \ and\ \bibinfo {author} {\bibfnamefont {A.}~\bibnamefont {Belmonte}},\
  }\href {\doibase 10.1103/physreve.93.032138} {\bibfield  {journal} {\bibinfo
  {journal} {Phys. Rev. E}\ }\textbf {\bibinfo {volume} {93}} (\bibinfo {year}
  {2016}),\ 10.1103/physreve.93.032138}\BibitemShut {NoStop}%
\bibitem [{\citenamefont {Blokhuis}\ \emph {et~al.}(2018)\citenamefont
  {Blokhuis}, \citenamefont {Lacoste}, \citenamefont {Nghe},\ and\
  \citenamefont {Peliti}}]{blokhuis2018prl}%
  \BibitemOpen
  \bibfield  {author} {\bibinfo {author} {\bibfnamefont {A.}~\bibnamefont
  {Blokhuis}}, \bibinfo {author} {\bibfnamefont {D.}~\bibnamefont {Lacoste}},
  \bibinfo {author} {\bibfnamefont {P.}~\bibnamefont {Nghe}}, \ and\ \bibinfo
  {author} {\bibfnamefont {L.}~\bibnamefont {Peliti}},\ }\href {\doibase
  10.1103/physrevlett.120.158101} {\bibfield  {journal} {\bibinfo  {journal}
  {Phys. Rev. Lett.}\ }\textbf {\bibinfo {volume} {120}} (\bibinfo {year}
  {2018}),\ 10.1103/physrevlett.120.158101}\BibitemShut {NoStop}%
\bibitem [{\citenamefont {Jiang}\ \emph {et~al.}(2018)\citenamefont {Jiang},
  \citenamefont {Chen}, \citenamefont {Huang},\ and\ \citenamefont
  {Lai}}]{jiang2018pre}%
  \BibitemOpen
  \bibfield  {author} {\bibinfo {author} {\bibfnamefont {J.}~\bibnamefont
  {Jiang}}, \bibinfo {author} {\bibfnamefont {Y.-Z.}\ \bibnamefont {Chen}},
  \bibinfo {author} {\bibfnamefont {Z.-G.}\ \bibnamefont {Huang}}, \ and\
  \bibinfo {author} {\bibfnamefont {Y.-C.}\ \bibnamefont {Lai}},\ }\href
  {\doibase 10.1103/physreve.98.042305} {\bibfield  {journal} {\bibinfo
  {journal} {Phys. Rev. E}\ }\textbf {\bibinfo {volume} {98}} (\bibinfo {year}
  {2018}),\ 10.1103/physreve.98.042305}\BibitemShut {NoStop}%
\bibitem [{\citenamefont {Artiges}\ \emph {et~al.}(2019)\citenamefont
  {Artiges}, \citenamefont {Gracia-L{\'{a}}zaro}, \citenamefont
  {Flor{\'{\i}}a},\ and\ \citenamefont {Moreno}}]{artiges2019pre}%
  \BibitemOpen
  \bibfield  {author} {\bibinfo {author} {\bibfnamefont {E.}~\bibnamefont
  {Artiges}}, \bibinfo {author} {\bibfnamefont {C.}~\bibnamefont
  {Gracia-L{\'{a}}zaro}}, \bibinfo {author} {\bibfnamefont {L.~M.}\
  \bibnamefont {Flor{\'{\i}}a}}, \ and\ \bibinfo {author} {\bibfnamefont
  {Y.}~\bibnamefont {Moreno}},\ }\href {\doibase 10.1103/physreve.100.052307}
  {\bibfield  {journal} {\bibinfo  {journal} {Phys. Rev. E}\ }\textbf {\bibinfo
  {volume} {100}} (\bibinfo {year} {2019}),\
  10.1103/physreve.100.052307}\BibitemShut {NoStop}%
\bibitem [{\citenamefont {Allahverdyan}\ \emph {et~al.}(2019)\citenamefont
  {Allahverdyan}, \citenamefont {Babajanyan},\ and\ \citenamefont
  {Hu}}]{allahverdyan2019pre}%
  \BibitemOpen
  \bibfield  {author} {\bibinfo {author} {\bibfnamefont {A.~E.}\ \bibnamefont
  {Allahverdyan}}, \bibinfo {author} {\bibfnamefont {S.~G.}\ \bibnamefont
  {Babajanyan}}, \ and\ \bibinfo {author} {\bibfnamefont {C.-K.}\ \bibnamefont
  {Hu}},\ }\href {\doibase 10.1103/physreve.100.032401} {\bibfield  {journal}
  {\bibinfo  {journal} {Phys. Rev. E}\ }\textbf {\bibinfo {volume} {100}}
  (\bibinfo {year} {2019}),\ 10.1103/physreve.100.032401}\BibitemShut {NoStop}%
\bibitem [{\citenamefont {Barfuss}\ \emph {et~al.}(2019)\citenamefont
  {Barfuss}, \citenamefont {Donges},\ and\ \citenamefont
  {Kurths}}]{barfuss2019pre}%
  \BibitemOpen
  \bibfield  {author} {\bibinfo {author} {\bibfnamefont {W.}~\bibnamefont
  {Barfuss}}, \bibinfo {author} {\bibfnamefont {J.~F.}\ \bibnamefont {Donges}},
  \ and\ \bibinfo {author} {\bibfnamefont {J.}~\bibnamefont {Kurths}},\ }\href
  {\doibase 10.1103/physreve.99.043305} {\bibfield  {journal} {\bibinfo
  {journal} {Phys. Rev. E}\ }\textbf {\bibinfo {volume} {99}} (\bibinfo {year}
  {2019}),\ 10.1103/physreve.99.043305}\BibitemShut {NoStop}%
\bibitem [{\citenamefont {Wang}\ \emph {et~al.}(2019)\citenamefont {Wang},
  \citenamefont {Gu}, \citenamefont {Zhao},\ and\ \citenamefont
  {Quan}}]{wang2019pre}%
  \BibitemOpen
  \bibfield  {author} {\bibinfo {author} {\bibfnamefont {X.}~\bibnamefont
  {Wang}}, \bibinfo {author} {\bibfnamefont {C.}~\bibnamefont {Gu}}, \bibinfo
  {author} {\bibfnamefont {J.}~\bibnamefont {Zhao}}, \ and\ \bibinfo {author}
  {\bibfnamefont {J.}~\bibnamefont {Quan}},\ }\href {\doibase
  10.1103/physreve.100.022411} {\bibfield  {journal} {\bibinfo  {journal}
  {Phys. Rev. E}\ }\textbf {\bibinfo {volume} {100}} (\bibinfo {year} {2019}),\
  10.1103/physreve.100.022411}\BibitemShut {NoStop}%
\bibitem [{\citenamefont {Yamamoto}\ \emph {et~al.}(2019)\citenamefont
  {Yamamoto}, \citenamefont {Okada}, \citenamefont {Taguchi},\ and\
  \citenamefont {Muto}}]{yamamoto2019pre}%
  \BibitemOpen
  \bibfield  {author} {\bibinfo {author} {\bibfnamefont {H.}~\bibnamefont
  {Yamamoto}}, \bibinfo {author} {\bibfnamefont {I.}~\bibnamefont {Okada}},
  \bibinfo {author} {\bibfnamefont {T.}~\bibnamefont {Taguchi}}, \ and\
  \bibinfo {author} {\bibfnamefont {M.}~\bibnamefont {Muto}},\ }\href {\doibase
  10.1103/physreve.100.032304} {\bibfield  {journal} {\bibinfo  {journal}
  {Phys. Rev. E}\ }\textbf {\bibinfo {volume} {100}} (\bibinfo {year} {2019}),\
  10.1103/physreve.100.032304}\BibitemShut {NoStop}%
\bibitem [{\citenamefont {Stepan}(2009)}]{stepan2009rsta}%
  \BibitemOpen
  \bibfield  {author} {\bibinfo {author} {\bibfnamefont {G.}~\bibnamefont
  {Stepan}},\ }\href {\doibase 10.1098/rsta.2008.0279} {\bibfield  {journal}
  {\bibinfo  {journal} {Philos. Trans. R. Soc. A}\ } (\bibinfo {year} {2009}),\
  10.1098/rsta.2008.0279}\BibitemShut {NoStop}%
\bibitem [{\citenamefont {Erneux}\ \emph {et~al.}(2017)\citenamefont {Erneux},
  \citenamefont {Javaloyes}, \citenamefont {Wolfrum},\ and\ \citenamefont
  {Yanchuk}}]{erneux2017chaos}%
  \BibitemOpen
  \bibfield  {author} {\bibinfo {author} {\bibfnamefont {T.}~\bibnamefont
  {Erneux}}, \bibinfo {author} {\bibfnamefont {J.}~\bibnamefont {Javaloyes}},
  \bibinfo {author} {\bibfnamefont {M.}~\bibnamefont {Wolfrum}}, \ and\
  \bibinfo {author} {\bibfnamefont {S.}~\bibnamefont {Yanchuk}},\ }\href
  {\doibase 10.1063/1.5011354} {\bibfield  {journal} {\bibinfo  {journal}
  {Chaos}\ }\textbf {\bibinfo {volume} {27}},\ \bibinfo {pages} {114201}
  (\bibinfo {year} {2017})}\BibitemShut {NoStop}%
\bibitem [{\citenamefont
  {Alonso-Sanz}(2009{\natexlab{a}})}]{alonso-sanz2009ijbc}%
  \BibitemOpen
  \bibfield  {author} {\bibinfo {author} {\bibfnamefont {R.}~\bibnamefont
  {Alonso-Sanz}},\ }\href {\doibase 10.1142/S0218127409024554} {\bibfield
  {journal} {\bibinfo  {journal} {Int. J. Bifurc. Chaos Appl. Sci. Eng.}\
  }\textbf {\bibinfo {volume} {19}},\ \bibinfo {pages} {2899} (\bibinfo {year}
  {2009}{\natexlab{a}})}\BibitemShut {NoStop}%
\bibitem [{\citenamefont
  {Alonso-Sanz}(2009{\natexlab{b}})}]{alonso-sanz2009chaos}%
  \BibitemOpen
  \bibfield  {author} {\bibinfo {author} {\bibfnamefont {R.}~\bibnamefont
  {Alonso-Sanz}},\ }\href {\doibase 10.1063/1.3106322} {\bibfield  {journal}
  {\bibinfo  {journal} {Chaos}\ }\textbf {\bibinfo {volume} {19}},\ \bibinfo
  {pages} {023102} (\bibinfo {year} {2009}{\natexlab{b}})}\BibitemShut
  {NoStop}%
\bibitem [{\citenamefont {Wang}\ \emph {et~al.}(2015)\citenamefont {Wang},
  \citenamefont {Chen}, \citenamefont {Yang}, \citenamefont {Zou},\ and\
  \citenamefont {Luo}}]{wang2015biosystems}%
  \BibitemOpen
  \bibfield  {author} {\bibinfo {author} {\bibfnamefont {T.}~\bibnamefont
  {Wang}}, \bibinfo {author} {\bibfnamefont {Z.}~\bibnamefont {Chen}}, \bibinfo
  {author} {\bibfnamefont {L.}~\bibnamefont {Yang}}, \bibinfo {author}
  {\bibfnamefont {Y.}~\bibnamefont {Zou}}, \ and\ \bibinfo {author}
  {\bibfnamefont {J.}~\bibnamefont {Luo}},\ }\href {\doibase
  https://doi.org/10.1016/j.biosystems.2015.03.006} {\bibfield  {journal}
  {\bibinfo  {journal} {Biosystems}\ }\textbf {\bibinfo {volume} {131}},\
  \bibinfo {pages} {30 } (\bibinfo {year} {2015})}\BibitemShut {NoStop}%
\bibitem [{\citenamefont {Yi}\ and\ \citenamefont {Zuwang}(1997)}]{yi1997jtb}%
  \BibitemOpen
  \bibfield  {author} {\bibinfo {author} {\bibfnamefont {T.}~\bibnamefont
  {Yi}}\ and\ \bibinfo {author} {\bibfnamefont {W.}~\bibnamefont {Zuwang}},\
  }\href {\doibase 10.1006/jtbi.1997.0427} {\bibfield  {journal} {\bibinfo
  {journal} {J. Theor. Biol.}\ }\textbf {\bibinfo {volume} {187}},\ \bibinfo
  {pages} {111} (\bibinfo {year} {1997})}\BibitemShut {NoStop}%
\bibitem [{\citenamefont {Alboszta}\ and\ \citenamefont
  {Mie{\c{}}kisz}(2004)}]{alboszta2004jtb}%
  \BibitemOpen
  \bibfield  {author} {\bibinfo {author} {\bibfnamefont {J.}~\bibnamefont
  {Alboszta}}\ and\ \bibinfo {author} {\bibfnamefont {J.}~\bibnamefont
  {Mie{\c{}}kisz}},\ }\href {\doibase 10.1016/j.jtbi.2004.06.012} {\bibfield
  {journal} {\bibinfo  {journal} {J. Theor. Biol.}\ }\textbf {\bibinfo {volume}
  {231}},\ \bibinfo {pages} {175} (\bibinfo {year} {2004})}\BibitemShut
  {NoStop}%
\bibitem [{\citenamefont {Wesson}\ \emph {et~al.}(2016)\citenamefont {Wesson},
  \citenamefont {Rand},\ and\ \citenamefont {Rand}}]{wesson2016ijbc}%
  \BibitemOpen
  \bibfield  {author} {\bibinfo {author} {\bibfnamefont {E.}~\bibnamefont
  {Wesson}}, \bibinfo {author} {\bibfnamefont {R.}~\bibnamefont {Rand}}, \ and\
  \bibinfo {author} {\bibfnamefont {D.}~\bibnamefont {Rand}},\ }\href {\doibase
  10.1142/s0218127416500061} {\bibfield  {journal} {\bibinfo  {journal} {Int.
  J. Bifurc. Chaos Appl. Sci. Eng.}\ }\textbf {\bibinfo {volume} {26}},\
  \bibinfo {pages} {1650006} (\bibinfo {year} {2016})}\BibitemShut {NoStop}%
\bibitem [{\citenamefont {Moreira}\ \emph {et~al.}(2012)\citenamefont
  {Moreira}, \citenamefont {Pinheiro}, \citenamefont {Nunes},\ and\
  \citenamefont {Pacheco}}]{moreira2012jtb}%
  \BibitemOpen
  \bibfield  {author} {\bibinfo {author} {\bibfnamefont {J.~A.}\ \bibnamefont
  {Moreira}}, \bibinfo {author} {\bibfnamefont {F.~L.}\ \bibnamefont
  {Pinheiro}}, \bibinfo {author} {\bibfnamefont {A.}~\bibnamefont {Nunes}}, \
  and\ \bibinfo {author} {\bibfnamefont {J.~M.}\ \bibnamefont {Pacheco}},\
  }\href {\doibase 10.1016/j.jtbi.2011.12.027} {\bibfield  {journal} {\bibinfo
  {journal} {J. Theor. Biol.}\ }\textbf {\bibinfo {volume} {298}},\ \bibinfo
  {pages} {8} (\bibinfo {year} {2012})}\BibitemShut {NoStop}%
\bibitem [{\citenamefont {Iijima}(2012)}]{ijima2012jtb}%
  \BibitemOpen
  \bibfield  {author} {\bibinfo {author} {\bibfnamefont {R.}~\bibnamefont
  {Iijima}},\ }\href {\doibase 10.1016/j.jtbi.2012.01.001} {\bibfield
  {journal} {\bibinfo  {journal} {J. Theor. Biol.}\ }\textbf {\bibinfo {volume}
  {300}},\ \bibinfo {pages} {1} (\bibinfo {year} {2012})}\BibitemShut {NoStop}%
\bibitem [{\citenamefont {Iijima}(2011)}]{ijima2011mss}%
  \BibitemOpen
  \bibfield  {author} {\bibinfo {author} {\bibfnamefont {R.}~\bibnamefont
  {Iijima}},\ }\href {\doibase 10.1016/j.mathsocsci.2010.12.002} {\bibfield
  {journal} {\bibinfo  {journal} {Math. Soc. Sci.}\ }\textbf {\bibinfo {volume}
  {61}},\ \bibinfo {pages} {83} (\bibinfo {year} {2011})}\BibitemShut {NoStop}%
\bibitem [{\citenamefont {Burridge}\ \emph {et~al.}(2017)\citenamefont
  {Burridge}, \citenamefont {Gao},\ and\ \citenamefont
  {Mao}}]{burridge2017epjb}%
  \BibitemOpen
  \bibfield  {author} {\bibinfo {author} {\bibfnamefont {J.}~\bibnamefont
  {Burridge}}, \bibinfo {author} {\bibfnamefont {Y.}~\bibnamefont {Gao}}, \
  and\ \bibinfo {author} {\bibfnamefont {Y.}~\bibnamefont {Mao}},\ }\href
  {\doibase 10.1140/epjb/e2016-70471-1} {\bibfield  {journal} {\bibinfo
  {journal} {Eur. Phys. J. B}\ }\textbf {\bibinfo {volume} {90}} (\bibinfo
  {year} {2017}),\ 10.1140/epjb/e2016-70471-1}\BibitemShut {NoStop}%
\bibitem [{\citenamefont {Toupo}\ and\ \citenamefont
  {Strogatz}(2015)}]{toupo2015pre}%
  \BibitemOpen
  \bibfield  {author} {\bibinfo {author} {\bibfnamefont {D.~F.~P.}\
  \bibnamefont {Toupo}}\ and\ \bibinfo {author} {\bibfnamefont {S.~H.}\
  \bibnamefont {Strogatz}},\ }\href {\doibase 10.1103/physreve.91.052907}
  {\bibfield  {journal} {\bibinfo  {journal} {Phys. Rev. E}\ }\textbf {\bibinfo
  {volume} {91}} (\bibinfo {year} {2015}),\
  10.1103/physreve.91.052907}\BibitemShut {NoStop}%
\bibitem [{\citenamefont {Komarova}\ \emph {et~al.}(2001)\citenamefont
  {Komarova}, \citenamefont {Niyogi},\ and\ \citenamefont
  {Nowak}}]{komarova2001jtb}%
  \BibitemOpen
  \bibfield  {author} {\bibinfo {author} {\bibfnamefont {N.~L.}\ \bibnamefont
  {Komarova}}, \bibinfo {author} {\bibfnamefont {P.}~\bibnamefont {Niyogi}}, \
  and\ \bibinfo {author} {\bibfnamefont {M.~A.}\ \bibnamefont {Nowak}},\ }\href
  {\doibase 10.1006/jtbi.2000.2240} {\bibfield  {journal} {\bibinfo  {journal}
  {J. Theor. Biol.}\ }\textbf {\bibinfo {volume} {209}},\ \bibinfo {pages} {43}
  (\bibinfo {year} {2001})}\BibitemShut {NoStop}%
\bibitem [{\citenamefont {Mobilia}(2010)}]{mobilia2010jtb}%
  \BibitemOpen
  \bibfield  {author} {\bibinfo {author} {\bibfnamefont {M.}~\bibnamefont
  {Mobilia}},\ }\href {\doibase 10.1016/j.jtbi.2010.01.008} {\bibfield
  {journal} {\bibinfo  {journal} {J. Theor. Biol.}\ }\textbf {\bibinfo {volume}
  {264}},\ \bibinfo {pages} {1} (\bibinfo {year} {2010})}\BibitemShut {NoStop}%
\bibitem [{\citenamefont {Toupo}\ \emph {et~al.}(2014)\citenamefont {Toupo},
  \citenamefont {Rand},\ and\ \citenamefont {Strogatz}}]{toupo2014ijbc}%
  \BibitemOpen
  \bibfield  {author} {\bibinfo {author} {\bibfnamefont {D.~F.~P.}\
  \bibnamefont {Toupo}}, \bibinfo {author} {\bibfnamefont {D.~G.}\ \bibnamefont
  {Rand}}, \ and\ \bibinfo {author} {\bibfnamefont {S.~H.}\ \bibnamefont
  {Strogatz}},\ }\href {\doibase 10.1142/s0218127414300353} {\bibfield
  {journal} {\bibinfo  {journal} {Int. J. Bifurc. Chaos Appl. Sci. Eng.}\
  }\textbf {\bibinfo {volume} {24}},\ \bibinfo {pages} {1430035} (\bibinfo
  {year} {2014})}\BibitemShut {NoStop}%
\bibitem [{\citenamefont {Shalom}(1986)}]{shalom1986jp}%
  \BibitemOpen
  \bibfield  {author} {\bibinfo {author} {\bibfnamefont {S.~R.}\ \bibnamefont
  {Shalom}},\ }\href {http://www.jstor.org/stable/2130953} {\bibfield
  {journal} {\bibinfo  {journal} {J. Politics}\ }\textbf {\bibinfo {volume}
  {48}},\ \bibinfo {pages} {234} (\bibinfo {year} {1986})}\BibitemShut
  {NoStop}%
\bibitem [{\citenamefont {Hilbe}\ \emph {et~al.}(2018)\citenamefont {Hilbe},
  \citenamefont {{\v{S}}imsa}, \citenamefont {Chatterjee},\ and\ \citenamefont
  {Nowak}}]{hilbe2018nature}%
  \BibitemOpen
  \bibfield  {author} {\bibinfo {author} {\bibfnamefont {C.}~\bibnamefont
  {Hilbe}}, \bibinfo {author} {\bibfnamefont {{\v{S}}.}~\bibnamefont
  {{\v{S}}imsa}}, \bibinfo {author} {\bibfnamefont {K.}~\bibnamefont
  {Chatterjee}}, \ and\ \bibinfo {author} {\bibfnamefont {M.~A.}\ \bibnamefont
  {Nowak}},\ }\href {\doibase 10.1038/s41586-018-0277-x} {\bibfield  {journal}
  {\bibinfo  {journal} {Nature}\ }\textbf {\bibinfo {volume} {559}},\ \bibinfo
  {pages} {246} (\bibinfo {year} {2018})}\BibitemShut {NoStop}%
\bibitem [{\citenamefont {Nowak}\ and\ \citenamefont
  {May}(1992)}]{nowak1992nature}%
  \BibitemOpen
  \bibfield  {author} {\bibinfo {author} {\bibfnamefont {M.~A.}\ \bibnamefont
  {Nowak}}\ and\ \bibinfo {author} {\bibfnamefont {R.~M.}\ \bibnamefont
  {May}},\ }\href {\doibase 10.1038/359826a0} {\bibfield  {journal} {\bibinfo
  {journal} {Nature}\ }\textbf {\bibinfo {volume} {359}},\ \bibinfo {pages}
  {826} (\bibinfo {year} {1992})}\BibitemShut {NoStop}%
\bibitem [{\citenamefont {Fang}\ \emph {et~al.}(2002)\citenamefont {Fang},
  \citenamefont {Kimbrough}, \citenamefont {Pace}, \citenamefont {Valluri},\
  and\ \citenamefont {Zheng}}]{fang2002gdn}%
  \BibitemOpen
  \bibfield  {author} {\bibinfo {author} {\bibfnamefont {C.}~\bibnamefont
  {Fang}}, \bibinfo {author} {\bibfnamefont {S.~O.}\ \bibnamefont {Kimbrough}},
  \bibinfo {author} {\bibfnamefont {S.}~\bibnamefont {Pace}}, \bibinfo {author}
  {\bibfnamefont {A.}~\bibnamefont {Valluri}}, \ and\ \bibinfo {author}
  {\bibfnamefont {Z.}~\bibnamefont {Zheng}},\ }\href {\doibase
  10.1023/A:1020639132471} {\bibfield  {journal} {\bibinfo  {journal} {Group
  Decis. Negot.}\ }\textbf {\bibinfo {volume} {11}},\ \bibinfo {pages} {449}
  (\bibinfo {year} {2002})}\BibitemShut {NoStop}%
\bibitem [{\citenamefont {Doebeli}(2004)}]{doebeli2004science}%
  \BibitemOpen
  \bibfield  {author} {\bibinfo {author} {\bibfnamefont {M.}~\bibnamefont
  {Doebeli}},\ }\href {\doibase 10.1126/science.1101456} {\bibfield  {journal}
  {\bibinfo  {journal} {Science}\ }\textbf {\bibinfo {volume} {306}},\ \bibinfo
  {pages} {859} (\bibinfo {year} {2004})}\BibitemShut {NoStop}%
\bibitem [{\citenamefont {K\"{u}mmerli}\ \emph {et~al.}(2007)\citenamefont
  {K\"{u}mmerli}, \citenamefont {Colliard}, \citenamefont {Fiechter},
  \citenamefont {Petitpierre}, \citenamefont {Russier},\ and\ \citenamefont
  {Keller}}]{kummerli2007prsb}%
  \BibitemOpen
  \bibfield  {author} {\bibinfo {author} {\bibfnamefont {R.}~\bibnamefont
  {K\"{u}mmerli}}, \bibinfo {author} {\bibfnamefont {C.}~\bibnamefont
  {Colliard}}, \bibinfo {author} {\bibfnamefont {N.}~\bibnamefont {Fiechter}},
  \bibinfo {author} {\bibfnamefont {B.}~\bibnamefont {Petitpierre}}, \bibinfo
  {author} {\bibfnamefont {F.}~\bibnamefont {Russier}}, \ and\ \bibinfo
  {author} {\bibfnamefont {L.}~\bibnamefont {Keller}},\ }\href {\doibase
  10.1098/rspb.2007.0793} {\bibfield  {journal} {\bibinfo  {journal} {Proc. R.
  Soc. B}\ }\textbf {\bibinfo {volume} {274}},\ \bibinfo {pages} {2965}
  (\bibinfo {year} {2007})}\BibitemShut {NoStop}%
\bibitem [{\citenamefont {Barker}(2017)}]{Barker2017book}%
  \BibitemOpen
  \bibfield  {author} {\bibinfo {author} {\bibfnamefont {J.~L.}\ \bibnamefont
  {Barker}},\ }in\ \href {\doibase 10.1007/978-3-319-16999-6_1220-1} {\emph
  {\bibinfo {booktitle} {Encyclopedia of Evolutionary Psychological Science}}}\
  (\bibinfo  {publisher} {Springer International Publishing},\ \bibinfo {year}
  {2017})\ pp.\ \bibinfo {pages} {1--8}\BibitemShut {NoStop}%
\bibitem [{\citenamefont {Skyrms}(2001)}]{skyrms2001paapa}%
  \BibitemOpen
  \bibfield  {author} {\bibinfo {author} {\bibfnamefont {B.}~\bibnamefont
  {Skyrms}},\ }\href {\doibase 10.2307/3218711} {\bibfield  {journal} {\bibinfo
   {journal} {Proceedings and Addresses of the American Philosophical
  Association}\ }\textbf {\bibinfo {volume} {75}},\ \bibinfo {pages} {31}
  (\bibinfo {year} {2001})}\BibitemShut {NoStop}%
\bibitem [{\citenamefont {Tembine}\ \emph {et~al.}(2007)\citenamefont
  {Tembine}, \citenamefont {Altman},\ and\ \citenamefont
  {El-Azouzi}}]{tembine2007article}%
  \BibitemOpen
  \bibfield  {author} {\bibinfo {author} {\bibfnamefont {H.}~\bibnamefont
  {Tembine}}, \bibinfo {author} {\bibfnamefont {E.}~\bibnamefont {Altman}}, \
  and\ \bibinfo {author} {\bibfnamefont {R.}~\bibnamefont {El-Azouzi}},\ }in\
  \href {http://dl.acm.org/citation.cfm?id=1345263.1345309} {\emph {\bibinfo
  {booktitle} {Proceedings of the 2Nd International Conference on Performance
  Evaluation Methodologies and Tools}}},\ \bibinfo {series and number}
  {ValueTools '07}\ (\bibinfo  {publisher} {ICST (Institute for Computer
  Sciences, Social-Informatics and Telecommunications Engineering)},\ \bibinfo
  {address} {ICST, Brussels, Belgium, Belgium},\ \bibinfo {year} {2007})\ pp.\
  \bibinfo {pages} {36:1--36:8}\BibitemShut {NoStop}%
\bibitem [{\citenamefont {Roose}\ and\ \citenamefont
  {Szalai}(2007)}]{roose2007book}%
  \BibitemOpen
  \bibfield  {author} {\bibinfo {author} {\bibfnamefont {D.}~\bibnamefont
  {Roose}}\ and\ \bibinfo {author} {\bibfnamefont {R.}~\bibnamefont {Szalai}},\
  }in\ \href {\doibase 10.1007/978-1-4020-6356-5_12} {\emph {\bibinfo
  {booktitle} {Understanding Complex Systems}}}\ (\bibinfo  {publisher}
  {Springer Netherlands},\ \bibinfo {year} {2007})\ pp.\ \bibinfo {pages}
  {359--399}\BibitemShut {NoStop}%
\bibitem [{\citenamefont {Asl}\ and\ \citenamefont
  {Ulsoy}(2003)}]{asl2003jdsmc}%
  \BibitemOpen
  \bibfield  {author} {\bibinfo {author} {\bibfnamefont {F.~M.}\ \bibnamefont
  {Asl}}\ and\ \bibinfo {author} {\bibfnamefont {A.~G.}\ \bibnamefont
  {Ulsoy}},\ }\href {\doibase 10.1115/1.1568121} {\bibfield  {journal}
  {\bibinfo  {journal} {J. Dyn. Syst. Meas. Control.}\ }\textbf {\bibinfo
  {volume} {125}},\ \bibinfo {pages} {215} (\bibinfo {year}
  {2003})}\BibitemShut {NoStop}%
\bibitem [{\citenamefont {Shinozaki}\ and\ \citenamefont
  {Mori}(2006)}]{shinozaki2006automatica}%
  \BibitemOpen
  \bibfield  {author} {\bibinfo {author} {\bibfnamefont {H.}~\bibnamefont
  {Shinozaki}}\ and\ \bibinfo {author} {\bibfnamefont {T.}~\bibnamefont
  {Mori}},\ }\href {\doibase https://doi.org/10.1016/j.automatica.2006.05.008}
  {\bibfield  {journal} {\bibinfo  {journal} {Automatica}\ }\textbf {\bibinfo
  {volume} {42}},\ \bibinfo {pages} {1791 } (\bibinfo {year}
  {2006})}\BibitemShut {NoStop}%
\bibitem [{\citenamefont {Yu}\ \emph {et~al.}(2017)\citenamefont {Yu},
  \citenamefont {Guo}, \citenamefont {Wang},\ and\ \citenamefont
  {Yang}}]{yu2017jmci}%
  \BibitemOpen
  \bibfield  {author} {\bibinfo {author} {\bibfnamefont {H.}~\bibnamefont
  {Yu}}, \bibinfo {author} {\bibfnamefont {S.}~\bibnamefont {Guo}}, \bibinfo
  {author} {\bibfnamefont {F.}~\bibnamefont {Wang}}, \ and\ \bibinfo {author}
  {\bibfnamefont {Y.}~\bibnamefont {Yang}},\ }\href {\doibase
  10.1093/imamci/dnx011} {\bibfield  {journal} {\bibinfo  {journal} {IMA J.
  Math. Control Inf.}\ }\textbf {\bibinfo {volume} {35}},\ \bibinfo {pages}
  {1005} (\bibinfo {year} {2017})}\BibitemShut {NoStop}%
\bibitem [{\citenamefont {Yi}\ \emph {et~al.}(2007)\citenamefont {Yi},
  \citenamefont {Nelson},\ and\ \citenamefont {Ulsoy}}]{yi2007mbe}%
  \BibitemOpen
  \bibfield  {author} {\bibinfo {author} {\bibfnamefont {S.}~\bibnamefont
  {Yi}}, \bibinfo {author} {\bibfnamefont {P.~W.}\ \bibnamefont {Nelson}}, \
  and\ \bibinfo {author} {\bibfnamefont {A.~G.}\ \bibnamefont {Ulsoy}},\ }\href
  {\doibase 10.3934/mbe.2007.4.355} {\bibfield  {journal} {\bibinfo  {journal}
  {Math. Biosci.}\ }\textbf {\bibinfo {volume} {4}},\ \bibinfo {pages} {355}
  (\bibinfo {year} {2007})}\BibitemShut {NoStop}%
\bibitem [{\citenamefont {Lehtonen}(2016)}]{lehtonen2016mee}%
  \BibitemOpen
  \bibfield  {author} {\bibinfo {author} {\bibfnamefont {J.}~\bibnamefont
  {Lehtonen}},\ }\href {\doibase 10.1111/2041-210x.12568} {\bibfield  {journal}
  {\bibinfo  {journal} {Methods Ecol. Evol.}\ }\textbf {\bibinfo {volume}
  {7}},\ \bibinfo {pages} {1110} (\bibinfo {year} {2016})}\BibitemShut
  {NoStop}%
\bibitem [{\citenamefont {Bortz}(2015)}]{bortz2015ifac}%
  \BibitemOpen
  \bibfield  {author} {\bibinfo {author} {\bibfnamefont {D.~M.}\ \bibnamefont
  {Bortz}},\ }\href {\doibase 10.1016/j.ifacol.2015.09.345} {\bibfield
  {journal} {\bibinfo  {journal} {{IFAC}-{PapersOnLine}}\ }\textbf {\bibinfo
  {volume} {48}},\ \bibinfo {pages} {13} (\bibinfo {year} {2015})}\BibitemShut
  {NoStop}%
\bibitem [{\citenamefont {Kuang}(1993)}]{kuang1993book}%
  \BibitemOpen
  \bibfield  {author} {\bibinfo {author} {\bibfnamefont {Y.}~\bibnamefont
  {Kuang}},\ }\href {https://catalogue.nla.gov.au/Record/1173989} {\emph
  {\bibinfo {title} {Delay Differential Equation with Application in Population
  Dynamics}}},\ edited by\ \bibinfo {editor} {\bibfnamefont {W.~F.}\
  \bibnamefont {Ames}}\ (\bibinfo  {publisher} {Academic Press, 1993},\
  \bibinfo {year} {1993})\ p.\ \bibinfo {pages} {398}\BibitemShut {NoStop}%
\bibitem [{\citenamefont {Gopalsamy}(1992)}]{gopalsamy1992book}%
  \BibitemOpen
  \bibfield  {author} {\bibinfo {author} {\bibfnamefont {K.}~\bibnamefont
  {Gopalsamy}},\ }\href {\doibase 10.1007/978-94-015-7920-9} {\emph {\bibinfo
  {title} {Stability and Oscillations in Delay Differential Equations of
  Population Dynamics}}}\ (\bibinfo  {publisher} {Springer Netherlands},\
  \bibinfo {year} {1992})\BibitemShut {NoStop}%
\bibitem [{\citenamefont {Arino}\ \emph {et~al.}(2006)\citenamefont {Arino},
  \citenamefont {Hbid},\ and\ \citenamefont {Dads}}]{arino2006book}%
  \BibitemOpen
  \bibinfo {editor} {\bibfnamefont {O.}~\bibnamefont {Arino}}, \bibinfo
  {editor} {\bibfnamefont {M.}~\bibnamefont {Hbid}}, \ and\ \bibinfo {editor}
  {\bibfnamefont {E.~A.}\ \bibnamefont {Dads}},\ eds.,\ \href {\doibase
  10.1007/1-4020-3647-7} {\emph {\bibinfo {title} {Delay Differential Equations
  and Applications}}}\ (\bibinfo  {publisher} {Springer Netherlands},\ \bibinfo
  {year} {2006})\BibitemShut {NoStop}%
\bibitem [{\citenamefont {Yu}\ \emph {et~al.}(2016)\citenamefont {Yu},
  \citenamefont {Guo},\ and\ \citenamefont {Wang}}]{yu2016namc}%
  \BibitemOpen
  \bibfield  {author} {\bibinfo {author} {\bibfnamefont {H.}~\bibnamefont
  {Yu}}, \bibinfo {author} {\bibfnamefont {S.}~\bibnamefont {Guo}}, \ and\
  \bibinfo {author} {\bibfnamefont {F.}~\bibnamefont {Wang}},\ }\href {\doibase
  10.15388/NA.2016.4.4} {\bibfield  {journal} {\bibinfo  {journal} {Nonlinear
  Anal-Model}\ } (\bibinfo {year} {2016}),\ 10.15388/NA.2016.4.4}\BibitemShut
  {NoStop}%
\bibitem [{\citenamefont {Li}\ \emph {et~al.}(2008)\citenamefont {Li},
  \citenamefont {Yan},\ and\ \citenamefont {Zhang}}]{li2008csf}%
  \BibitemOpen
  \bibfield  {author} {\bibinfo {author} {\bibfnamefont {W.-T.}\ \bibnamefont
  {Li}}, \bibinfo {author} {\bibfnamefont {X.-P.}\ \bibnamefont {Yan}}, \ and\
  \bibinfo {author} {\bibfnamefont {C.-H.}\ \bibnamefont {Zhang}},\ }\href
  {\doibase 10.1016/j.chaos.2006.11.015} {\bibfield  {journal} {\bibinfo
  {journal} {Chaos Solitons Fractals}\ }\textbf {\bibinfo {volume} {38}},\
  \bibinfo {pages} {227} (\bibinfo {year} {2008})}\BibitemShut {NoStop}%
\bibitem [{\citenamefont {Pandit}\ \emph {et~al.}(2018)\citenamefont {Pandit},
  \citenamefont {Mukhopadhyay},\ and\ \citenamefont
  {Chakraborty}}]{pandit2018chaos}%
  \BibitemOpen
  \bibfield  {author} {\bibinfo {author} {\bibfnamefont {V.}~\bibnamefont
  {Pandit}}, \bibinfo {author} {\bibfnamefont {A.}~\bibnamefont
  {Mukhopadhyay}}, \ and\ \bibinfo {author} {\bibfnamefont {S.}~\bibnamefont
  {Chakraborty}},\ }\href {\doibase 10.1063/1.5011955} {\bibfield  {journal}
  {\bibinfo  {journal} {Chaos}\ }\textbf {\bibinfo {volume} {28}},\ \bibinfo
  {pages} {033104} (\bibinfo {year} {2018})}\BibitemShut {NoStop}%
\bibitem [{\citenamefont {Weissing}(1991)}]{weissing1991gem}%
  \BibitemOpen
  \bibfield  {author} {\bibinfo {author} {\bibfnamefont {F.~J.}\ \bibnamefont
  {Weissing}},\ }\enquote {\bibinfo {title} {Evolutionary stability and dynamic
  stability in a class of evolutionary normal form games},}\ in\ \href
  {\doibase 10.1007/978-3-662-02674-8_4} {\emph {\bibinfo {booktitle} {Game
  Equilibrium Models I: Evolution and Game Dynamics}}},\ \bibinfo {editor}
  {edited by\ \bibinfo {editor} {\bibfnamefont {R.}~\bibnamefont {Selten}}}\
  (\bibinfo  {publisher} {Springer Berlin Heidelberg},\ \bibinfo {address}
  {Berlin, Heidelberg},\ \bibinfo {year} {1991})\ pp.\ \bibinfo {pages}
  {29--97}\BibitemShut {NoStop}%
\bibitem [{\citenamefont {Hauert}(2002)}]{hauert2002ijbc}%
  \BibitemOpen
  \bibfield  {author} {\bibinfo {author} {\bibfnamefont {C.}~\bibnamefont
  {Hauert}},\ }\href {\doibase 10.1142/s0218127402005273} {\bibfield  {journal}
  {\bibinfo  {journal} {Int. J. Bifurc. Chaos Appl. Sci. Eng.}\ }\textbf
  {\bibinfo {volume} {12}},\ \bibinfo {pages} {1531} (\bibinfo {year}
  {2002})}\BibitemShut {NoStop}%
\bibitem [{\citenamefont {Hummert}\ \emph {et~al.}(2014)\citenamefont
  {Hummert}, \citenamefont {Bohl}, \citenamefont {Basanta}, \citenamefont
  {Deutsch}, \citenamefont {Werner}, \citenamefont {Thei{\ss}en}, \citenamefont
  {Schroeter},\ and\ \citenamefont {Schuster}}]{hummert2014mbs}%
  \BibitemOpen
  \bibfield  {author} {\bibinfo {author} {\bibfnamefont {S.}~\bibnamefont
  {Hummert}}, \bibinfo {author} {\bibfnamefont {K.}~\bibnamefont {Bohl}},
  \bibinfo {author} {\bibfnamefont {D.}~\bibnamefont {Basanta}}, \bibinfo
  {author} {\bibfnamefont {A.}~\bibnamefont {Deutsch}}, \bibinfo {author}
  {\bibfnamefont {S.}~\bibnamefont {Werner}}, \bibinfo {author} {\bibfnamefont
  {G.}~\bibnamefont {Thei{\ss}en}}, \bibinfo {author} {\bibfnamefont
  {A.}~\bibnamefont {Schroeter}}, \ and\ \bibinfo {author} {\bibfnamefont
  {S.}~\bibnamefont {Schuster}},\ }\href {\doibase 10.1039/c3mb70602h}
  {\bibfield  {journal} {\bibinfo  {journal} {Mol. {BioSyst}.}\ }\textbf
  {\bibinfo {volume} {10}},\ \bibinfo {pages} {3044} (\bibinfo {year}
  {2014})}\BibitemShut {NoStop}%
\end{thebibliography}%
\section*{APPENDICES}
\appendix
\section{Case of No Mutation}
\label{app:A}
\begin{figure}
\centering
\includegraphics[scale=0.4]{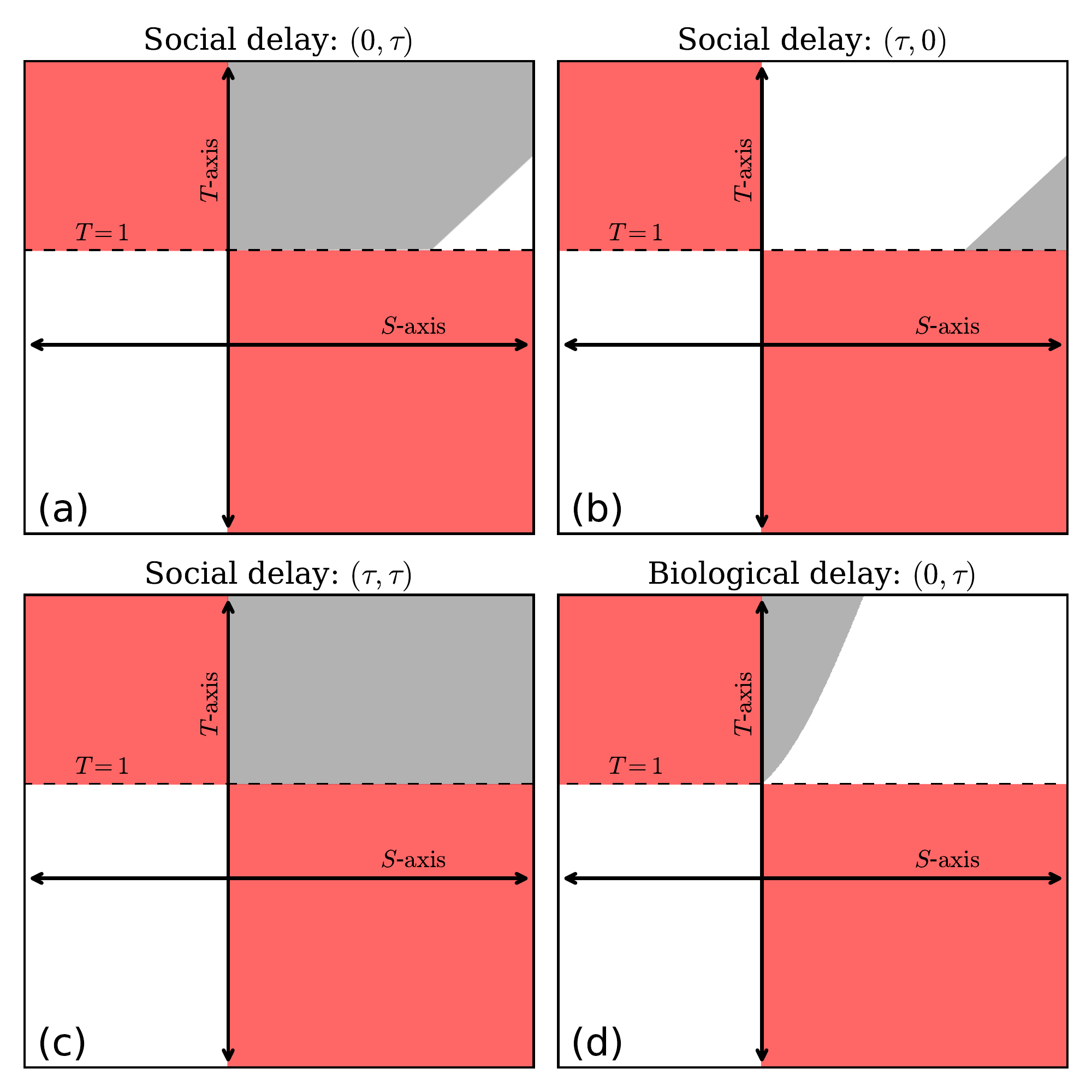}      	
\caption{\emph{(Colour online)} Emergence of limit cycle in the absence of mutation: The grey region schematically marks the region in $S$-$T$ space where the interior physical fixed point, $x_m$, can undergo the Hopf bifurcation for the cases of (a) social asymmetric delay $(0,\tau)$, (b) social asymmetric delay $(\tau,0)$, (c) social symmetric delay $(\tau,\tau)$, and (d) biological asymmetric delay $(0,\tau)$. Other cases are not shown as there is no possibility of a limit cycle in those cases. The red region emphasizes that no interior physical fixed point and hence, a physical limit cycle around it, is impossible.}	  
\label{fig:9}
\end{figure} 
In this appendix, we briefly present when stable limit cycles can emerge in two-play-two-person symmetric games under the replicator equation with social and biological delay, i.e., Eq.~(\ref{eq:social_explicit_form}) and Eq.~(\ref{eq:bio_explicit_form}) but with $q=1$ and $\mu=0$. It is known that there are actually twelve ordinally distinct games~\cite{hauert2002ijbc,hummert2014mbs,pandit2018chaos} whose payoff matrices, owing to positive affine transformations, can be compactly represented by a two-parameter matrix:
 \begin{eqnarray} 
 {\sf\Pi}=
\begin{tabular}{c}
 $\begin{bmatrix}  
1  & S   \\  T & 0 \\
\end{bmatrix}$;
\end{tabular}\; T,S \in {\mathbb{R}}.
\label{eq:PayOff_A1}
\end{eqnarray} 
 Since we are interested only in physical limit cycles, i.e., the limit cycles which do not go beyond the interval $[0,1]$, we should consider only those games that have a physical interior fixed point ($x_m$) that may undergo the Hopf bifurcation. This happens only for the SH class ($T<1$ and $S<0$) and the SD class ($T>1$ and $S>0$), each of which consists of three ordinally inequivalent games. Subsequently, following the method outlined in the main text of this paper, we find when the games within these classes \textcolor{black}{undergo} the Hopf bifurcation and a stable limit cycle emerges about $x_m$ when delay is above some threshold value. Fig.~\ref{fig:9} exhibits the conclusions for all the kinds of delays employed in this paper. We remark that our results regarding the emergence of stable limit cycles for social symmetric delay (refer Fig.~\ref{fig:9}c) matches with that found in literature~\citep{wesson2016ijbc}; other results are unreported elsewhere. Also, note that there is no limit cycle behaviour possible for the SH game in the absence of mutation.
\section{Ordinal Inequivalent SD Games}
\label{app:B}
\begin{figure}
\centering
\includegraphics[scale=0.4]{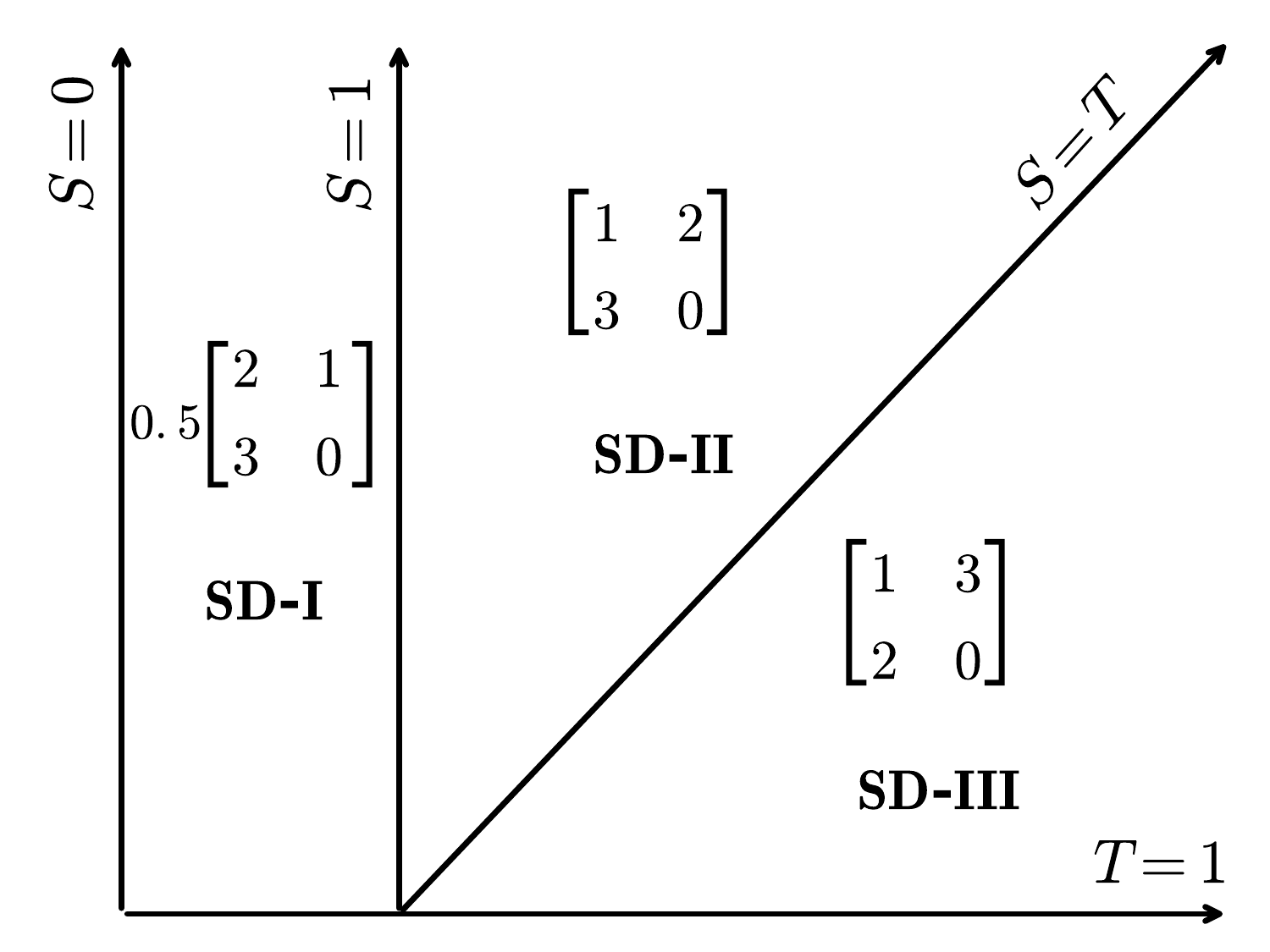}      		  
\caption{Three ordinally inequivalent SD games: The lines $S=0$, $T=1$, $S=1$, and $S=T$ divide the SD class of games in three distinct games, SD-I, SD-II, and SD-III. We show a representative payoff matrix for each of the games.}
\label{fig:10}
\end{figure}
\begin{figure}
\centering
\includegraphics[scale=0.42]{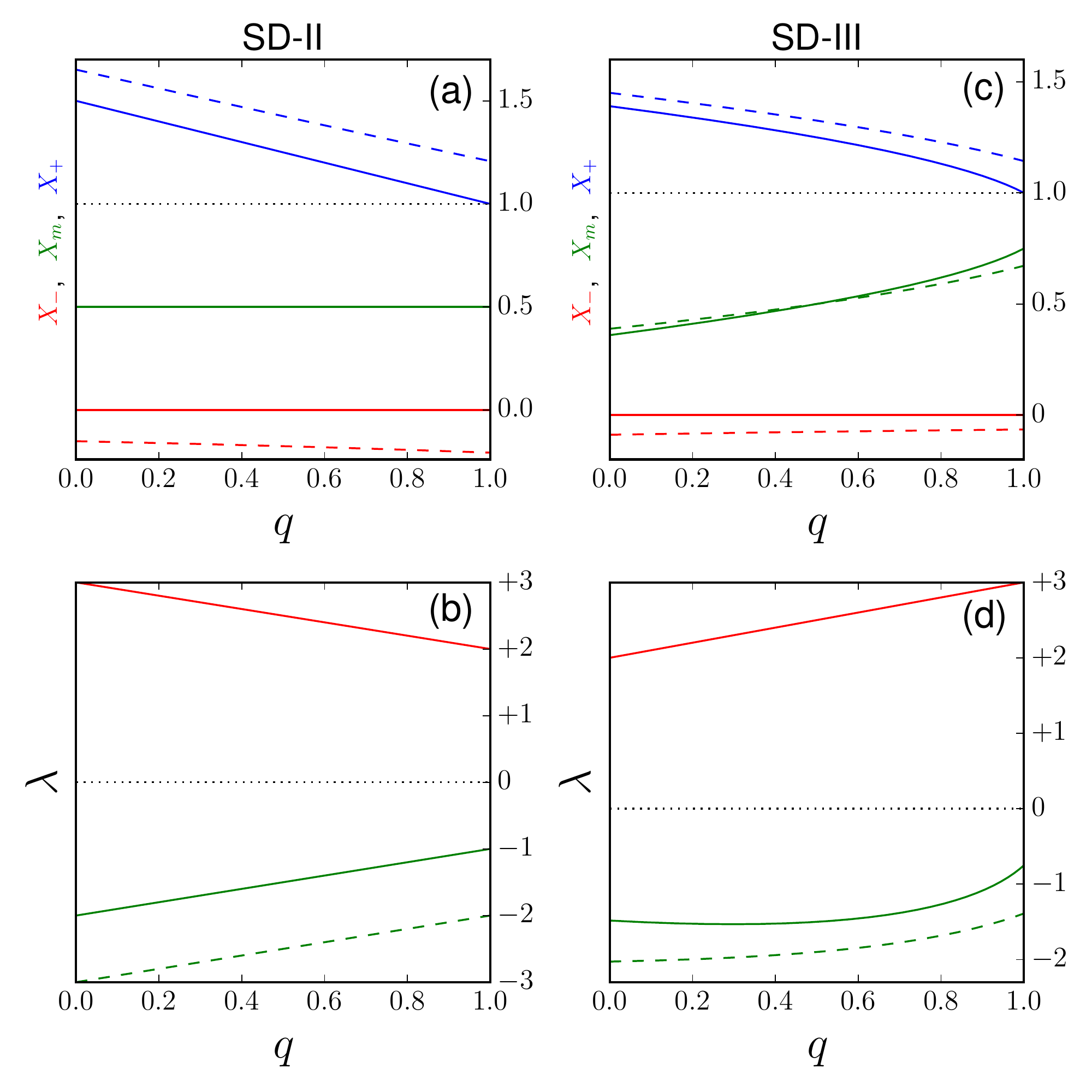}      		  
\caption{\emph{(Colour online)} Linear stability of the fixed points of the replicator-mutator equation corresponding to SD-II and SD-III. \textcolor{black}{The} top row depicts the variation of fixed points $X_m$ (green), $X_-$ (red) and $X_+$ (blue) with the change in mutation parameter $q$ (for two values of parameter $\mu$) when the payoff matrix corresponds to (a) SD-II game and (c) SD-III game. The corresponding eigenvalue, $\lambda$, of the Jacobian found in the course of linear stability analysis is plotted in the bottom row (following the colour conventions used for the top row) for (b) SD-II game and (d) SD-III game. \textcolor{black}{The} solid line stands for $\mu=0$ while the dashed line is for a non-zero $\mu$, specifically, $0.5$ for SD-II game and $0.2$ for SD-III game.}
\label{fig:11}
\end{figure}
\begin{figure}
\centering
\includegraphics[scale=0.4]{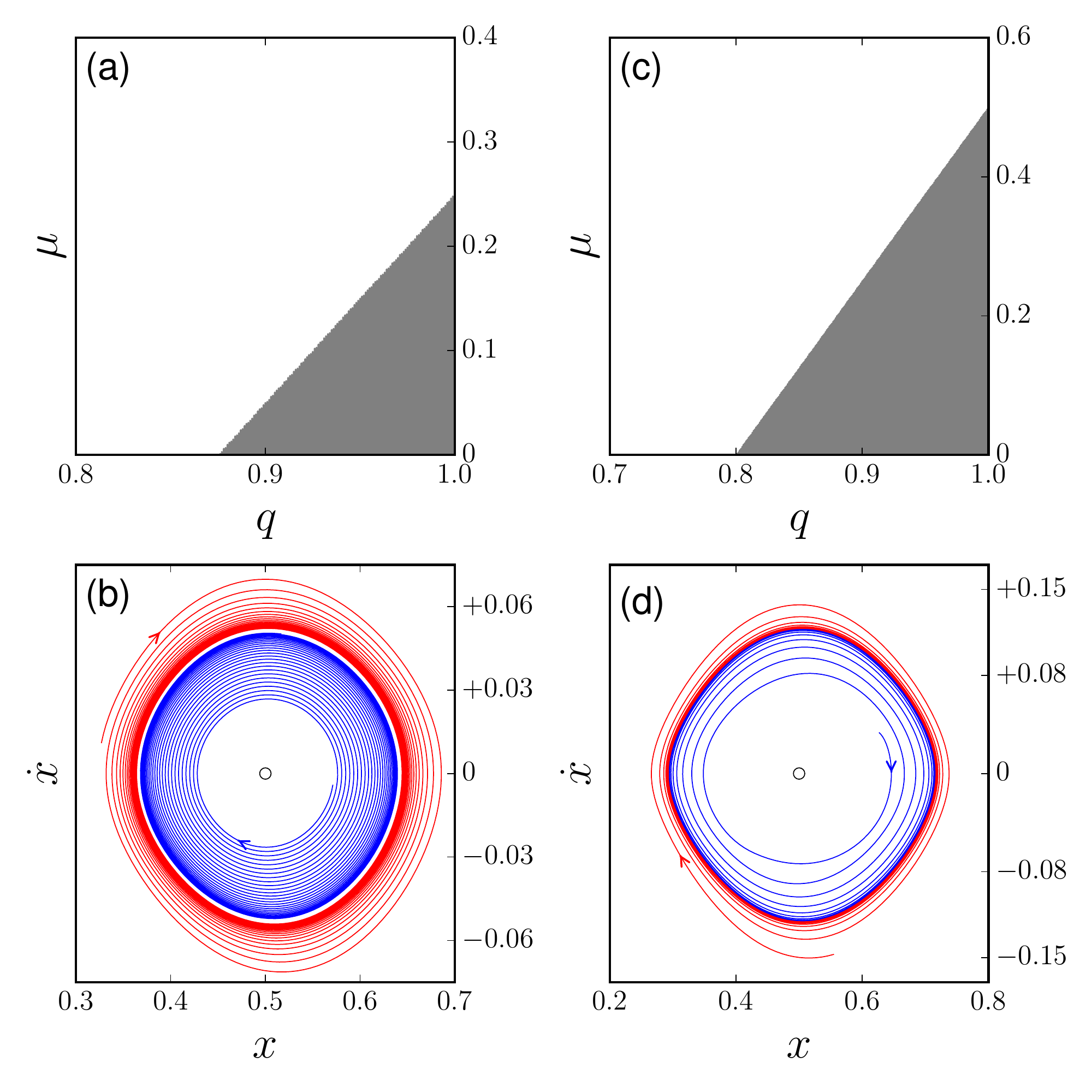}      		  
\caption{\emph{(Colour online)} Stable limit cycle emerges following the Hopf bifurcation in SD-II game for social delay: Grey areas in subplot (a) and (c) mark the regions in the mutation parameter space where stable limit cycle emerges at some threshold value of asymmetric delay $(0,\tau)$ and symmetric delay $(\tau,\tau)$ respectively. The fixed point that undergoes the Hopf bifurcation is $X_m$. Subplot (b) and (d) respectively showcases illustrative phase diagrams corresponding to $q=0.92$, $\mu=0.02$, and $(\tau_1,\tau_2)=(0,\tau)=(0,7)$; and  $q=0.9$, $\mu=0.1$ and $(\tau_1,\tau_2)=(\tau,\tau)=(5,5)$ picked from the grey regions. The unfilled circle represent unstable focus, $X_m$; and the blue and the red curves are representative phase trajectories approaching the limit cycles from inside and outside respectively.}
\label{fig:12}
\end{figure}
\begin{figure}
\centering
\includegraphics[scale=0.4]{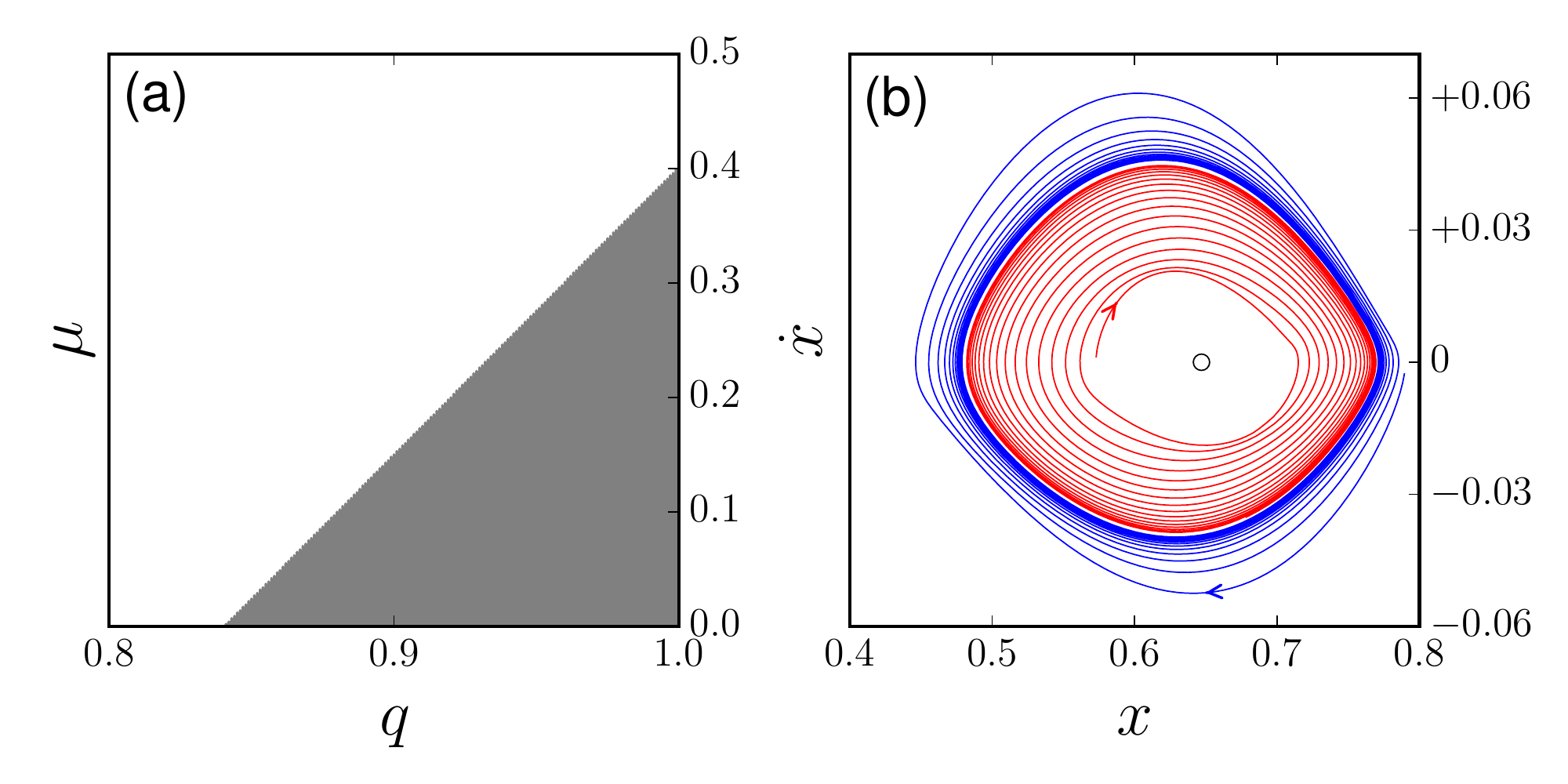}      		  
\caption{\emph{(Colour online)} Stable limit cycle emerges following the Hopf bifurcation in SD-III game for social symmetric delay $(\tau,\tau)$: Grey area in subplot (a) marks the region in the mutation parameter space where stable limit cycle emerges at some threshold value of delay. The fixed point that undergoes the Hopf bifurcation is $X_m$. Subplot (b) showcases an illustrative phase diagram corresponding to $q=0.9$, $\mu=0.1$, and $\tau=12$ picked from the grey region. The unfilled circle represents unstable focus, $X_m$; and the red and the blue curves are representative phase trajectories approaching the limit cycle from inside and outside respectively.}
\label{fig:12a}
\end{figure}
Since based on what the Nash equilibria are, two-player-two-strategy symmetric games can be classified into twelve ordinal classes, we compare the delayed replicator dynamics of the inequivalent ordinal games. To this end, the class of the SD games that has three ordinally distinct games as shown in Fig.~\ref{fig:10}. The representative payoff matrix that we have studied earlier within the SD class (refer Fig.~\ref{fig:1}) belongs to SD-I ordinal structure. To make our study complete, we study the dynamics of SD-II and SD-III ordinal class of games in the delayed replicator-mutator models.

In Fig.~\ref{fig:11} we showcase the variation of fixed points and their corresponding eigenvalues (on doing linear stability analysis) for the non-delayed case as a function of multiplicative mutation ($q$)  for two given additive mutation ($\mu$) values. We note that qualitatively the behaviours of SD-II and SD-III are same as that of SD-I: In \textcolor{black}{the} presence of mutation, the fixed point $X_+$ is unphysical (\textcolor{black}{a} value outside $[0,1]$). The other fixed point $X_-$ is unphysical whenever $\mu \ne 0$ (irrespective of the value of $q$). When $\mu=0$, $X_-=0$ for all possible values of $q$. $X_m$ is the only fixed point that is always existent irrespective of how much mutation is in action. Furthermore, as done for SD-I game earlier, we find the region in mutation parameter space where \textcolor{black}{a} stable limit cycle emerges following Hopf bifurcation (refer Fig.~\ref{fig:12}) for SD-II and SD-III ordinal games. We also illustrate the limit cycles using the phase plots corresponding to one point from the corresponding mutation parameter space in Fig.~\ref{fig:12}. It is interesting to observe that stable limit cycle emerges in SD-II for both social asymmetric and symmetric delay (Fig.~\ref{fig:12}a and Fig.~\ref{fig:12}c respectively), whereas SD-III shows stable limit cycle only for social symmetric delay (Fig.~\ref{fig:12a}a). 
\section{Cardinal Equivalent SD Games}
\label{app:C}
\begin{figure}
\centering
\includegraphics[scale=0.4]{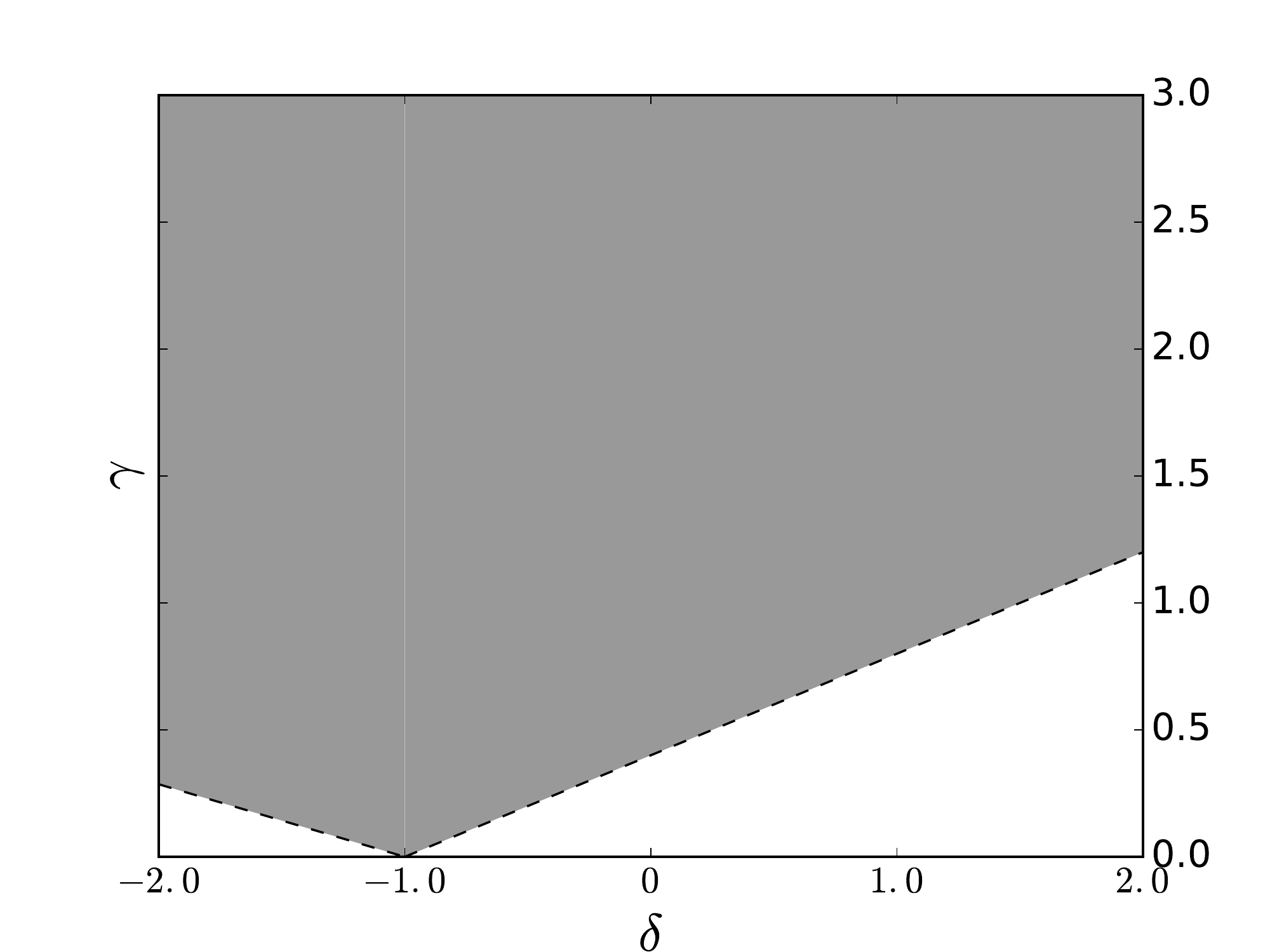}      		  
\caption{Emergence of limit cycle depends on scaling and shifting of the payoff matrix: We fix $q=0.9$ and $\mu=0.1$, and grey the region in $\delta$-$\gamma$ space (see  Eq.~(\ref{eq:PayOff_A4})) where $X_m$ undergoes the Hopf bifurcation for SD-I game with social asymmetric delay, $(0,\tau)$.}
\label{fig:13}
\end{figure}
\begin{figure*}
\centering
\includegraphics[scale=0.35]{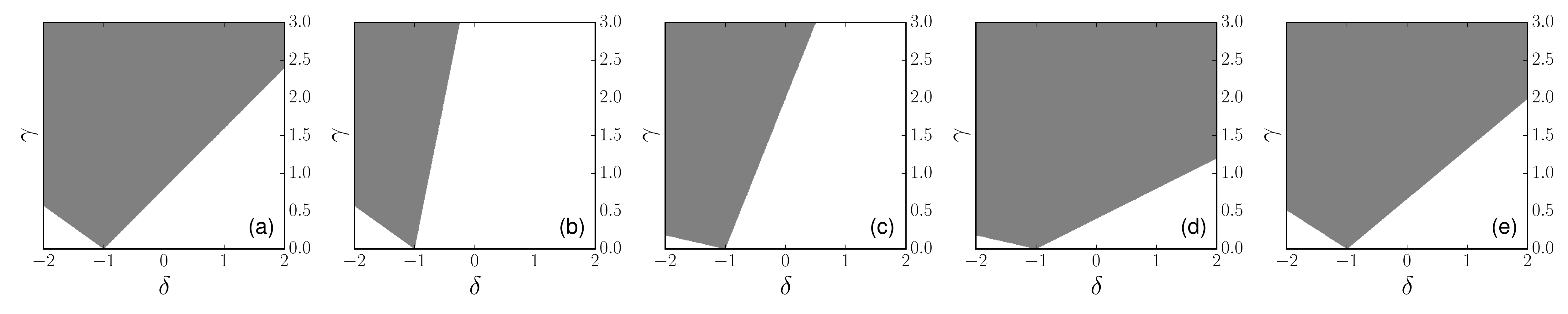}      		  
\caption{\textcolor{black}{The} emergence of \textcolor{black}{a} limit cycle depends on scaling and shifting of the payoff matrix: We fix $q=0.9$ and $\mu=0.1$, and grey the region in $\delta$-$\gamma$ space where $X_m$ undergoes the Hopf bifurcation for the SD games with payoff matrices given in Fig.~\ref{fig:10}. In particular, subplots (a)-(e) respectively correspond to the case of SD-I game with social asymmetric delay, $(0,\tau)$; SD-I game with social symmetric delay, $(\tau,\tau)$; SD-II game with social asymmetric delay, $(0,\tau)$; (d) SD-II game with social symmetric delay, $(\tau,\tau)$; and SD-III game with social symmetric delay, $(\tau,\tau)$.}
\label{fig:14}
\end{figure*}
We know that positive affine transformations, 
\begin{eqnarray} 
\begin{tabular}{c}
$\gamma$ $\begin{bmatrix}  
a  & b   \\ c & d \\

\end{bmatrix}$ +$\delta$

 $\begin{bmatrix}  
1  & 1   \\  1 & 1 \\
\end{bmatrix}$
=
 $\begin{bmatrix}  
a'  & b'   \\  c' & d' \\
\end{bmatrix}$;\;
\end{tabular}\vspace{1mm}\gamma \in {\mathbb{R}}^+, \delta \in {\mathbb{R}},\quad
\end{eqnarray} 
leads to cardinally equivalent games within a distinct ordinal structure of the payoff matrix. \textcolor{black}{The} same ordinal structure implies \textcolor{black}{the} same rational outcome for \textcolor{black}{a given one-shot} game. However, it doesn't guarantee that the solutions of the delayed replicator-mutator equation based on the payoff matrix is independent of the parameters, $\gamma$ and $\delta$, of the affine transformation. Hence, it is important to understand what significant change happens in the dynamics when one deals with cardinally equivalent games. For the sake of convenience and continuity, we yet again consider the SD class of games to find the region in $\gamma$-$\delta$ space where \textcolor{black}{a} stable limit cycle emerges following the Hopf bifurcation for a given mutation strength. 

Our line of argument to find stable limit cycle is \textcolor{black}{the} same as discussed in the main text. First, we consider SD-I game and use the payoff matrix given in Fig.~\ref{fig:1}). The matrix is scaled and shifted through the following positive affine transformation:
 \begin{eqnarray}
   {\sf \Pi} = 
\begin{tabular}{c}
$\gamma$ 
$\begin{bmatrix}  
2  & 1   \\  3 & 0 \\
\end{bmatrix}$
 +$\delta$
 $\begin{bmatrix}  
1  & 1   \\  1 & 1 \\
\end{bmatrix}$;\;
\end{tabular}\vspace{1mm}\gamma \in {\mathbb{R}}^+, \delta \in {\mathbb{R}}.
\label{eq:PayOff_A4}
\end{eqnarray}
The fixed points of the replicator-mutator dynamics for this cardinally transformed game when $q=0.9$ and $\mu=0.1$ are
\begin{eqnarray}
 &&X_m = \frac{1}{2}\,, \nonumber\\
&&X_- = \frac{1}{10\gamma} \left(6\gamma-\sqrt{36\gamma^2+10\delta\gamma+10\gamma}\right),\nonumber\\
&&X_+ = \frac{1}{10\gamma} \left(6\gamma+\sqrt{36\gamma^2+10\delta\gamma+10\gamma}\right).\nonumber \label{eqn3:FP_cardinal games}
\end{eqnarray}
The values of the mutation parameters are so chosen because they do give rise to stable limit cycles for social delay before any transformation is effected. On doing linear stability analyses about the fixed points and finding the characteristic equations having same form as in Eq.~(\ref{eq:characteristic_eq_3}), e.g, about $X_m$, we get $\alpha=6\gamma/10$ and $\beta = (-1/10)(\gamma+2\delta+2)$. Now, we find the region in $\gamma$-$\delta$ space where the Hopf bifurcation leads to the emergence of stable limit cycle (see Fig.~\ref{fig:13}) by imposing the conditions $\alpha=\max\{0,\beta\}$ and $\alpha+\beta>0$. We do similar calculation using the payoff matrices for SD-II and SD-III given in Fig.~\ref{fig:10}. We show the corresponding results in Fig.~\ref{fig:14} where only those cases of delay are shown where there is any possibility of stable limit cycle. In conclusion, there is no denying the fact that detailed dynamics depends on the exact values of the payoff matrix elements although there is a qualitative similarity in the results that the limit cycles, if at all, are still seen only for the social delay case in the SD game irrespective of whether it is SD-I, SD-II or SD-III. However, it must be kept in mind that we have not explored what happens for all possible payoff matrices that is a daunting task but it does not appear to us to be leading to any fundamentally new insight.

\end{document}